\documentclass[useAMS,usenatbib]{mn2e}
\usepackage[pdftex]{graphicx}
\usepackage{txfonts}

\usepackage{epstopdf}
\usepackage{url}
\usepackage{subfig}
\usepackage{longtable,lscape}

\usepackage[amssymb,mediumspace,thickqspace]{SIunits}
\addunit{\mol}{mol}
\addunit{\erg}{erg}
\addunit{\s}{s}
\addunit{\cm}{cm}
\addunit{\gramm}{g}

\pdfoutput=1

\def\arcsec{\hbox{$^{\prime\prime}$}}

\newcommand{\um}{\,\micro\metre}

\newcommand{\msun}{M_{\sun}}
\newcommand{\mbh}{M_\mathrm{BH}}

\newcommand{\lbol}{L_\mathrm{bol}}
\newcommand{\ledd}{\lambda_\mathrm{Edd}}
\newcommand{\Ledd}{L_\mathrm{Edd}}

\newcommand{\nh}{N_\mathrm{H}}

\newcommand{\lxi}{L^\mathrm{int}(\textrm{2-10\,keV})}

\newcommand{\ltw}{L(12\um)}

\newcommand{\rsub}{r_\mathrm{sub}}

\newcommand{\extnuc}{R_\textrm{AGN}^\textrm{ext}}
\newcommand{\polmid}{R_\textrm{MIDI}^\textrm{pol}}
\newcommand{\polagn}{R_\textrm{AGN}^\textrm{pol}}
\newcommand{\dmir}{d_\textrm{MIR}}
\newcommand{\dmirp}{d_\textrm{MIR}/\textrm{pc}}
\newcommand{\dmira}{d_\textrm{MIR}/\arcsec}
\newcommand{\dmirr}{d_\textrm{MIR}/r_\mathrm{sub}}

\newcommand{\tauK}{\tau_\mathrm{K}}
\newcommand{\pK}{p_\mathrm{K}}

\newcommand{\spitzerr}{{\it Spitzer}\ }                 % w/ space behind
\newcommand{\spitzer}{{\it Spitzer}}                    % w/o space behind

\newcommand{\isoo}{{\it ISO}\ }

\newcommand{\jwstt}{{\it JWST}\ }
\newcommand{\jwst}{{\it JWST}}

\newcommand{\oiv}{[O\,\textsc{IV}]\ }
\newcommand{\oiii}{[O\,\textsc{III}]\ }
\newcommand{\foiv}{F(\textrm{[O\,\textsc{IV}]})}

 \title[New evidence for polar dust]{New evidence for the ubiquity of prominent polar dust emission in AGN on tens of parsec scales} 

   \author[D. Asmus et al.]{D.~Asmus$^{1}$\thanks{E-mail: d.asmus@soton.ac.uk}\\
             $^1$Department of Physics \& Astronomy, University of Southampton, Hampshire SO17 1BJ, Southampton, United Kingdom\\
             }

\pagerange{\pageref{firstpage}--\pageref{lastpage}} \pubyear{2015}
\begin{document}

\label{firstpage}

\maketitle
 
\begin{abstract}
The key ingredient of active galactic nuclei (AGN) unification, the dusty obscuring torus was so far held responsible for the observed mid-infrared (MIR) emission of AGN.
However, the best studied objects with VLTI/MIDI show that instead a polar dusty wind is dominating these wavelengths, leaving little room for a torus contribution.
But is this wind an ubiquitous part of the AGN? 
To test this, we conducted a straightforward detection experiment, using the upgraded VLT/VISIR for deep subarcsecond resolution MIR imaging of a sample of nine [O\,\textsc{IV}]-bright, obscured AGN, all of which were predicted to have detectable polar emission.
Indeed, the new data reveal such emission in all objects but one.
We further estimate lower limits on the extent of the polar dust and show that the polar dust emission is dominating the total MIR emission of the AGN.
These findings support the scenario that polar dust is not only ubiquitous in AGN but also an integral part of its structure, processing a significant part of the primary radiation.
The polar dust has to be optically thin on average, which explains, e.g., the small dispersion in the observed mid-infrared--X-ray luminosity correlation.
At the same time, it has to be taken into account when deriving covering factors of obscuring material from mid-infrared to bolometric luminosity ratios.
Finally, we find a new tentative trend of increasing MIR emission size with increasing Eddington ratio.
\end{abstract}


%
%________________________________________________________________

\section{Introduction}
It seems beyond doubt that active galactic nuclei (AGN) contain large amounts of dust which is heated by the accretion onto the supermassive black hole, leading to copious amounts of mid-infrared (MIR) emission, which in fact contains roughly half of the total power output of these systems. 
It also is established that the innermost region of AGN is highly obscured for a significant fraction of their sky as originally inferred the presence of scattered light from the central region, visible in polarized light, and the presence of collimated ionization cones \citep{antonucci_unified_1993}.
This led to the common assumption that this obscuring material is distributed into a torus-like structure, which then would be also main MIR emitter,  (e.g., \citealt{nenkova_agn_2008}; see \citealt{netzer_revisiting_2015} and \citealt{almeida_nuclear_2017} for recent reviews).
However, this could never been directly shown, and there is growing evidence to the contrary.
Namely, the highest angular resolution MIR observations available obtained with VLTI/MIDI interferometry show that most of the MIR emission is coming from a polar extended component \citep{honig_parsec-scale_2012, honig_dust_2013, tristram_dusty_2014, lopez-gonzaga_revealing_2014, lopez-gonzaga_mid-infrared_2016, leftley_new_2018}.
It turned out that this polar component can also be resolved  with subarcsecond resolution single dish imaging in the MIR and, thus, extends up to scales of tens to hundreds of parsec
(e.g., \citealt{bock_high_2000, radomski_resolved_2003, packham_extended_2005, asmus_subarcsecond_2014, asmus_subarcsecond_2016}; hereafter A16).
Recently, it has been detected on even larger scales (and longer wavelengths; \citealt{fuller_sofia/forcast_2019}).
This means that this polar dust is present at similar scales as the narrow-line emitting region (NLR) of the AGN, and is probably forming a hollow cone-like structure surrounding the NLR.
In addition to the polar extended component, the VLTI/MIDI observations found a second, more compact component, which could be resolved into a equatorial disk in the objects with the best u,v coverage.
The two components can be interpreted as a dusty hollow cone, or hyperboloid, and a dusty thin disk as extension of the outer accretion disk, respectively.
Such a dust geometry can indeed successfully reproduce many observed infrared properties of the Seyferts  \citep{honig_dusty_2017}.
It also allowed us to explain all observed infrared features of the intrinsically best resolved AGN, in the Circinus galaxy at once \citep{stalevski_dissecting_2017, vollmer_thick_2018, stalevski_dissecting_2019}.
On the other hand, no match could be found with the classical clumpy torus model for Circinus.
In particular the observed morphology and interferometric visibilities could not reproduced, even after adding dust in the polar region.
 
Dynamically, \cite{honig_parsec-scale_2012} interpreted this polar dust structure as a dusty wind driven by the radiation of the accretion disk.
In that case, the actual obscurer could be a much more compact and quite hot structure like a puffed up outer accretion disk (see also \citealt{baskin_dust_2018}), as also assumed to occur in disks around forming stars, or a combination of such a puff-up and the base of the polar wind (see also \citealt{elitzur_agn-obscuring_2006} and \citealt{wada_radiation-driven_2012} for similar ideas).
In the best fitting model for Circinus by \cite{stalevski_dissecting_2019}, the base of the polar dusty wind is indeed optically thick and significantly contributes to the obscuration.
This raises the possibility that the obscuration along our line of sight to some Seyferts is dominated by the polar dust but we expect this to be the case  only for a small fraction owing  the polar dust component covering only $\sim 10\%$ of the sky as seen by the central source.
Then, on scales larger than a few parsec, the polar dusty wind becomes on average optically thin as it is required by the fact that we have a relatively clear line of sight towards the ionization cone.

From the physical side, the above scenario is also well motivated as radiation pressure has been suspected as major driver for the geometrical thickness of the obscuring material for a long time \citep{pier_radiation-pressure-supported_1992, konigl_disk-driven_1994, elvis_structure_2000, roth_three-dimensional_2012}.
In fact, polar dusty winds are now also successfully produced by hydrodynamical models \citep{chan_radiation-driven_2016,wada_multi-phase_2016, williamson_3d_2019}.

If this new scenario of the dust structure would hold true for the AGN population in general, it could have fundamental impact on the current AGN research field, which based many results on the presence of a infrared-prominent clumpy dust torus like, e.g., the determination of covering factors of obscuring material (e.g., \citealt{ramos_almeida_infrared_2009, ramos_almeida_testing_2011, alonso-herrero_torus_2011}).
Therefore, it is crucial to test how ubiquitous this polar dust phenomenon is.
One of the major caveats of the scenario is that it is based on the findings for relatively few objects so far.

A16 made the first statistical investigation of the polar dust phenomenon by looking at all archival subarcsecond resolution MIR images of those AGN which, on the one hand, show no nuclear starburst, and, on the other hand, are at least moderately powerful, i.e. no LINERs. 
Only a small fraction of the sources (21 of 149) showed robust evidence for extended emission.
This low detection rate was explained by the combination of several factors.
First, only in sufficiently inclined systems, i.e. obscured AGN, one expects to be able to clearly detect the polar emission.
Second, the objects have to be close and powerful enough, so that we can actually resolve the polar emission with direct imaging. 
And finally, most of the available data are relatively shallow, barely enough to detect the nuclear component in many cases, and thus insufficient to robustly detect extended emission.
So the results so far are consistent with every AGN having strong polar dust emission but its low detection rate leaves a lot of room for ambiguity.

To further test the ubiquity of polar dust emission, we have now performed a detection experiment by utilizing a prediction that is made in A16.
Namely, it was found that the amount of extended MIR emission correlates with the \oiv  25.89\,$\mu$m emission line flux.
This line originates from the ionization cone and, thus, comparable scales than the polar MIR emission. 
Thanks to in general negligible obscuration and resulting isotropy, the \oiv 25.89\,$\mu$m line has been found to be one of the best intrinsic indicators for the AGN bolometric luminosity \citep{melendez_new_2008}. 
Therefore, it is expected that objects with higher \oiv fluxes have either more powerful AGN, and thus probably larger cones, or are simply more nearby. 
Either way, a higher \oiv flux implies a larger apparent size on the sky.
This explains why all objects above a certain threshold in \oiv line flux, i.e., $\foiv \ge 6 \cdot 10^{-13}\,$erg/s/cm$^{-2}$, could be resolved and show polar MIR emission in the A16 sample.
The value of this threshold depends on our instrumentation.
%In other words, object below this threshold probably also have polar emission but we can can not detect it with current instrumentation.
Conclusively, the prediction is that all AGN with an \oiv flux larger than this threshold should have polar MIR emission detectable with 8 meter class telescopes.

\section{Sample Selection}\label{sec:sam}
To test the above prediction, we searched the NASA/IPAC Extragalactic Database (NED\footnote{http://ned.ipac.caltech.edu/}) for all galaxies with available \oiv measurements taken either with the \isoo or \spitzerr satellite \citep{kessler_infrared_1996,werner_spitzer_2004}
We found 596 local galaxies ($z\le0.1$), of which 32 have \oiv fluxes above the empirical polar dust detection threshold from A16 of $6 \cdot 10^{-13}\,$erg/s/cm$^{-2}$ ($\log \foiv/[\erg/\s] \ge -12.2$).
From those, we select only those 25 objects that harbour obscured AGN because for those the probability of being highly-inclined with respect to our line of sight is higher.
For these objects, the polar dust cones should lead to a clearly elongated MIR structure in the observations.
Furthermore, we exclude Mrk\,463E because this is a starbursting host in the merging process, leading to confusing MIR emission, and the very nearby starburst, but uncertain AGN, in NGC\,253 (e.g., \citealt{gunthardt_uncovering_2015}).
Of the remaining 23, 12 have already verified polar MIR emission.
This leaves 11 obscured AGN of which 8 are visible from Paranal.
Their basic properties including the \oiv fluxes are listed in Table~\ref{tab_sel}.

The prediction is that all of these 8 should exhibit detectable polar MIR emission.
To the sample, we further added another source, NGC\,2110, because previous MIR imaging indicated that its nucleus is possibly extended \citep{asmus_subarcsecond_2014}.
%{\bf
%Thanks to the \oiv being a robust intrinsic AGN power indicator, this sample should be relatively little biased with respect to the other AGN properties, making it well-suited to test our prediction representative for the local AGN population.
%}

% - histogram plot showing the selection, marking the selected and the already confirmed polar extended. --> maybe not
% - table with all objects and oiv fluxes and references in the appendix? --> rather not.

\begin{table}
\caption{Selected sample for new VISIR observations.}
\label{tab_sel}
\centering
\begin{tabular}{l c c c c}
\hline\hline
 & Opt. &  & $\log F$ & \\
Object & class & D & (\oiv) & Ref.\\
 &  & [Mpc] & [erg/cm/s$^{2}$] & \\
(1) & (2) & (3) & (4) & (5)\\
\hline
3C\,321 & 2.0 & 460.0 & -12.1 & 1\\
IC\,4518W & 2.0 & 76.1 & -12.1 & 2\\
Mrk\,573 & 2.0 & 73.1 & -12.1 & 3\\
NGC\,1365 & 1.8 & 17.9 & -11.8 & 4\\
NGC\,2110 & 2.0 & 35.9 & -12.3 & 5\\
NGC\,5135 & 2.0 & 66.0 & -12.2 & 6\\
NGC\,5506 & 1.9 & 31.6 & -11.7 & 4\\
NGC\,5643 & 2.0 & 20.9 & -12.1 & 4\\
NGC\,7582 & 1.8 & 23.0 & -11.6 & 6\\
\hline	
\end{tabular}
		                              
\begin{minipage}{1.0\columnwidth}
%\small
%\scriptsize
\normalsize
{\it -- Notes:} 
(1), (2), and (3) object name, optical class, and distance (D) from \cite{asmus_subarcsecond_2014};
(4) and (5) observed \oiv flux, $\foiv$, from \spitzer/IRS and corresponding reference:
1: \cite{dicken_spitzer_2014};
2: \cite{pereira-santaella_mid-infrared_2010};
3: \cite{sturm_mid-infrared_2002};
4: \cite{diamond-stanic_isotropic_2009};
5: \cite{weaver_mid-infrared_2010};
6: \cite{tommasin_spitzer-irs_2010};

\end{minipage}
\end{table}

\section{Observations}\label{sec:obs}
We observed the 9 selected obscured AGN  with the Very Large Telescope (VLT) mounted  Spectrometer and Imager for the Mid-infrared (VISIR; \citealt{lagage_successful_2004})  after its upgrade \citep{ kaufl_return_2015, kerber_visir_2016} in service mode between May 2017 and July 2018 (programme 099.B-0044; PI: Asmus).
Images were recorded  in the two filters B12.4 ($12.47 \pm0.5\,\um$) 
and Q1 ($17.65 \pm0.44\,\um$) with standard chopping and perpendicular nodding using a chop/nod throw of $8\,\arcsec$ and an on-source exposure times of 30\,min each.
For 3C\,321 and NGC\,5506, no successful observation in B12.4 could be made before the termination of the programme.
%The Q1 observations of NGC\,7582 was affected by technical problems and thus is excluded here.
To allow for accurate point-spread-function (PSF) subtraction, a nearby (within $10\degree$)  calibrator star taken from \cite{cohen_spectral_1999} was observed either directly before or after with 2\,min on-source exposure times per filter.
The data were custom reduced and analysed with an in-house developed, \texttt{python}-based pipeline.
Details of the observations and the individual measurements are listed in Table~\ref{tab_obs}.

\begin{table*}
\caption{Observations}
\label{tab_obs}
\centering
\begin{tabular}{l c c c c c c c c c c c}
\hline\hline
 &  &  & Cal. &  & Cal. & Cal. & Cal. & Sci. & Sci. & Sci. & Gauss\\
Object & Filter & Date & name & Sensit. & Maj. & Min. & PA & Maj. & Min. & PA & flux\\
 &  & [yyyy-mm-dd] &  & [mJy $10\sigma$/1h] & [as] & [as] & [$\degree$] & [as] & [as] & [$\degree$] & [mJy]\\
(1) & (2) & (3) & (4) & (5) & (6) & (7) & (8) & (9) & (10) & (11) & (12)\\
\hline
3C 321 & Q1 & 2018-07-10 & HD141992 & 25.5 & 0.49 & 0.45 & 4 & 1.03 & 0.68 & 99 & 154\\
IC 4518W & B12.4 & 2018-04-23 & HD136422 & 8.6 & 0.34 & 0.33 & 151 & 0.41 & 0.36 & 17 & 210\\
IC 4518W & Q1 & 2017-07-28 & HD136422 & 27.9 & 0.49 & 0.47 & 166 & 0.55 & 0.50 & 0 & 366\\
Mrk 573 & B12.4 & 2017-08-08 & HD10380 & 5.2 & 0.36 & 0.32 & 6 & 0.42 & 0.38 & 169 & 275\\
Mrk 573 & Q1 & 2017-08-06 & HD10380 & 22.8 & 0.45 & 0.43 & 0 & 0.50 & 0.45 & 157 & 461\\
NGC 1365 & B12.4 & 2017-08-30 & HD26967 & 5.9 & 0.37 & 0.36 & 63 & 0.43 & 0.38 & 102 & 434\\
NGC 1365 & Q1 & 2017-09-12 & HD26967 & 20.5 & 0.53 & 0.53 & 107 & 0.65 & 0.57 & 105 & 692\\
NGC 2110 & B12.4 & 2017-10-01 & HD39853 & 5.1 & 0.37 & 0.37 & 54 & 0.58 & 0.53 & 17 & 361\\
NGC 2110 & Q1 & 2017-09-27 & HD39853 & 26.0 & 0.46 & 0.44 & 17 & 0.47 & 0.46 & 165 & 569\\
NGC 5135 & B12.4 & 2018-07-23 & HD123139 & 10.4 & 0.42 & 0.39 & 6 & 0.43 & 0.38 & 1 & 123\\
NGC 5135 & Q1 & 2018-07-09 & HD123139 & 21.0 & 0.46 & 0.45 & 7 & 0.48 & 0.43 & 20 & 201\\
NGC 5506 & Q1 & 2018-07-08 & HD124294 & 26.1 & 0.45 & 0.45 & 19 & 0.49 & 0.46 & 24 & 1923\\
NGC 5643 & B12.4 & 2018-06-12 & HD136422 & 8.3 & 0.34 & 0.33 & 5 & 0.41 & 0.39 & 67 & 265\\
NGC 5643 & Q1 & 2017-07-26 & HD136422 & 25.0 & 0.46 & 0.44 & 8 & 0.50 & 0.49 & 55 & 855\\
NGC 7582 & B12.4 & 2017-09-04 & HD2261 & 9.0 & 0.38 & 0.35 & 130 & 0.38 & 0.38 & 82 & 494\\
NGC 7582 & Q1 & 2017-06-21 & HD2261 & 25.3 & 0.49 & 0.46 & 144 & 0.50 & 0.47 & 125 & 552\\
\hline	
\end{tabular}

\begin{minipage}{1.0\textwidth}
%\small
%\scriptsize
\normalsize
{\it -- Notes:} 
(1) object name;
(2) instrument filter B12.4 ($12.47 \pm0.5\,\um$) 
and Q1 ($17.65 \pm0.44\,\um$); 
(3) date of the observation;
(4) name of the calibrator star used as flux and PSF reference;
(5) derived sensitivity from the calibrator observation following the definition used by ESO (VISIR manual);
(6), (7) and (8) major and minor axis FWHM and corresponding position angle of the PSF derived from Gaussian fitting of the calibrator star;
(9), (10) and (11) major and minor axis FWHM and corresponding position angle of the PSF derived from Gaussian fitting of the science target;
(12) derived flux density from the Gaussian fitting of the science target; the associated uncertainty of this measurement is dominated by the systematic uncertainty on the flux of the calibrator star and is $10\%$.
\end{minipage}
\end{table*}

\section{Results \& Discussion}\label{sec:res}
\subsection{Presence of extended MIR emission}
\begin{figure*}
%    \centering
%    \sidecaption
   \includegraphics[angle=0,width=18cm]{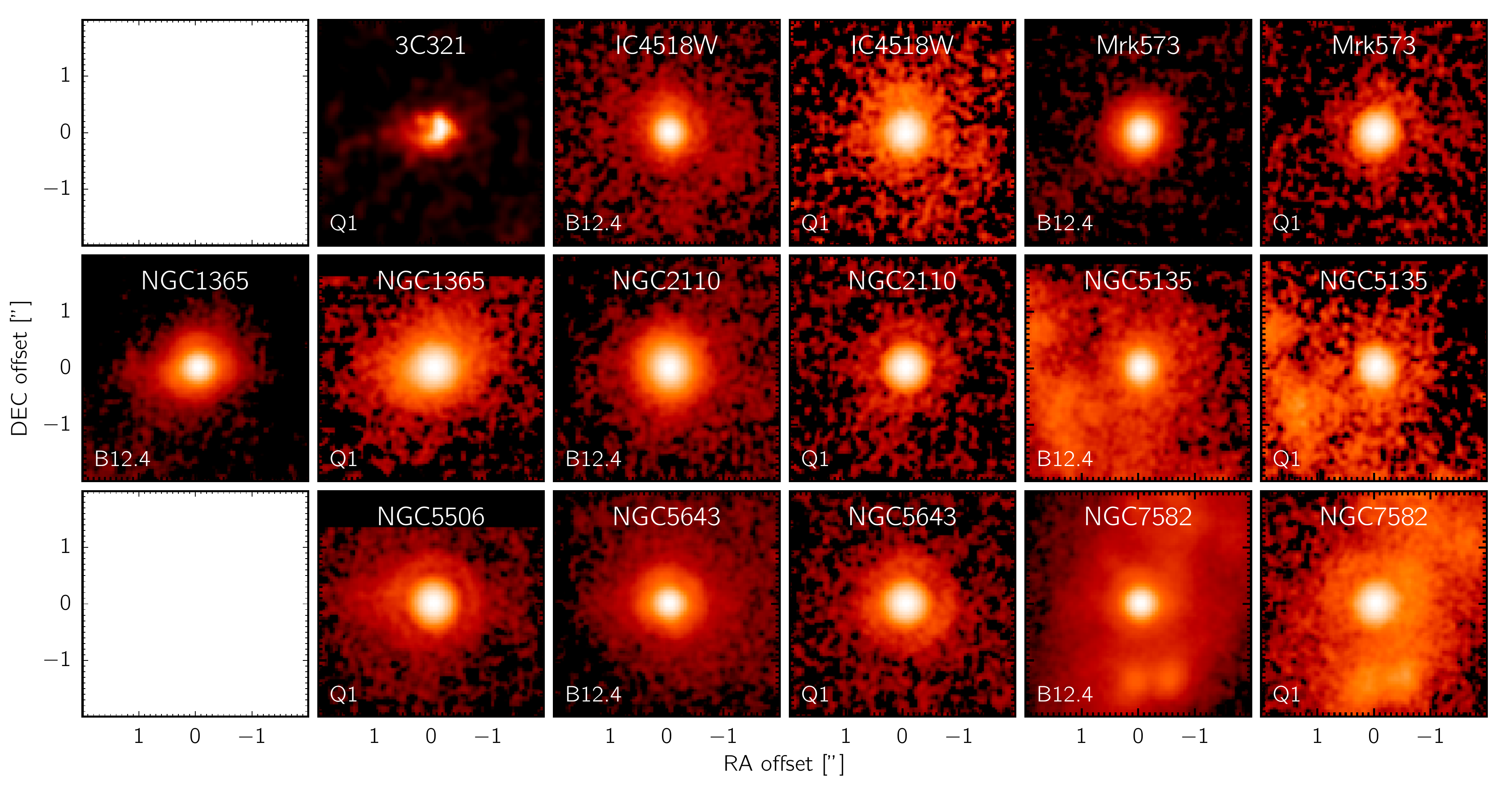}
    \caption{
             VISIR B12.4 ($12.47 \pm0.5\,\um$) 
and Q1 ($17.65 \pm0.44\,\um$) images of the central $4\arcsec \times 4\arcsec$ region of the observed AGN.
             The colour scaling is logarithmic in all images except for 3C\,321 (where it is linear) with black corresponding to the background level of the image and white to the brightest pixel. 
             All images where slightly smoothed with a Gaussian kernel with $\sigma=1\,$px.
            }
   \label{fig:gal}
\end{figure*}

The VISIR images obtained for all sources are shown in Fig.~\ref{fig:gal}.
All nuclei were clearly detected, whereas for NGC\,5135 and NGC\,7582 also the known kiloparsec-scale circum-nuclear starburst rings are partly visible (see, e.g.,  \citealt{asmus_subarcsecond_2014} for discussion). 
To roughly characterise the nuclear structures, we measure the total flux, extension and position angle (PA) of the nuclear emission with Gaussian fitting. 
The resulting values are listed in Table~\ref{tab_obs} in comparison to the similarly obtained values of the PSF reference stars.

All nuclei, except those of NGC\,5135 and NGC\,7582, show Gaussian-fit major axis full width half maximum (FWHM) values that are 10 to 20\% larger than those of the PSF references (median: 12\%).
In general, the level of resolvedness is 10\% larger in the B12.4 filter compared to that in the Q1 filter.
This is expected given the $\sim 30\%$ better angular resolution (0.34\arcsec versus 0.45\arcsec), and the $\sim3$ times better sensitivity in B12.4 compared to Q1 (7.5\,mJy versus 24\,mJy, respectively, for $10\sigma$ in one hour on-source integration time). 
In our case, these two advantages seem to outweigh the expected larger extent of the MIR emission at longer wavelengths as a result the centrally peak temperature distribution of the dust.
The nucleus of NGC\,1365 is here the only exception with the resolvedness being larger in Q1 compared to B12.4 ($23\%$ versus $16\%$).

\begin{figure*}
%    \centering
%   \sidecaption
   \includegraphics[angle=0,width=12cm]{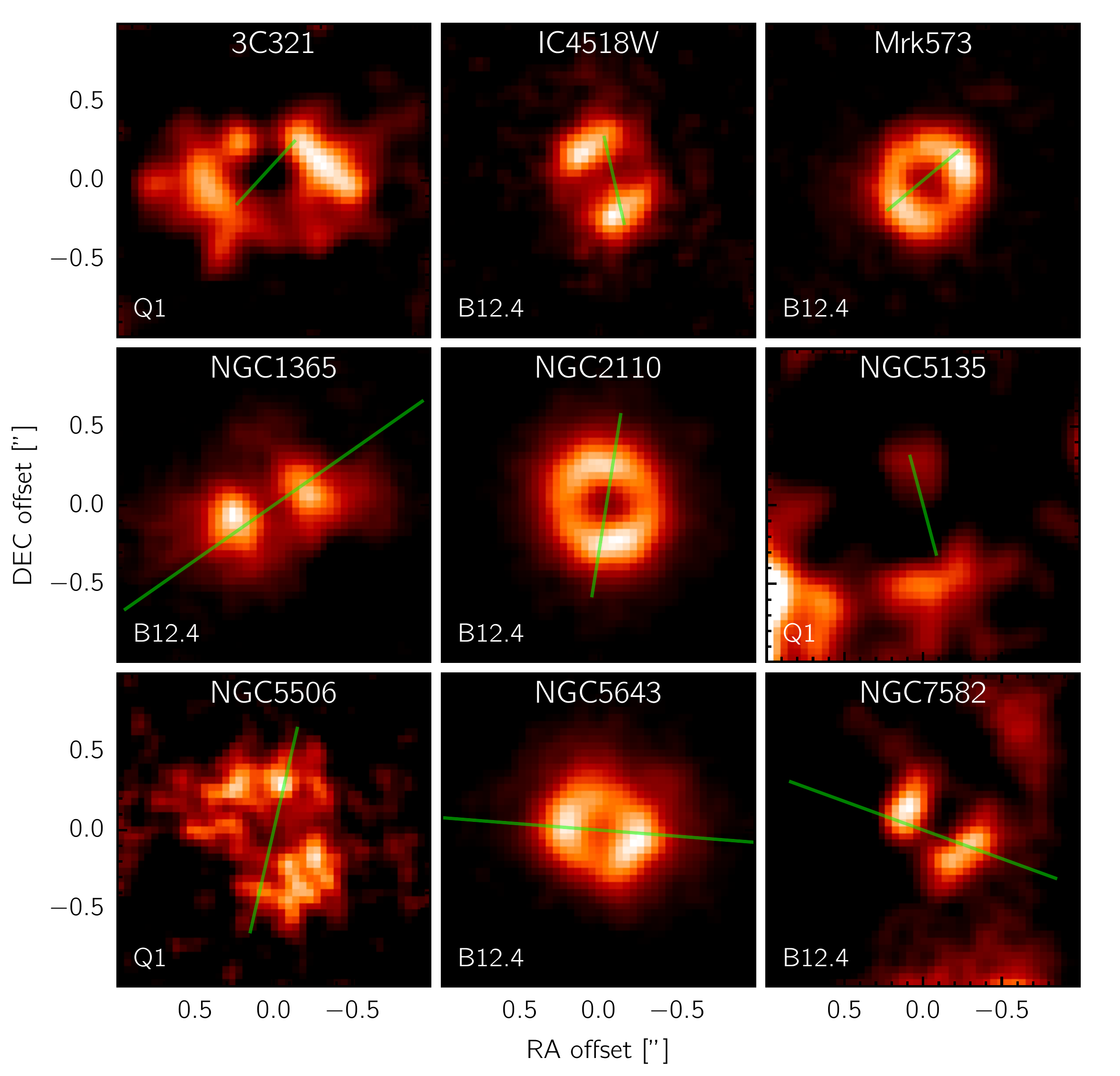}
    \caption{
             Zoom into the central $2\arcsec \times 2\arcsec$ region after deliberately over-subtracting the central point source with the corresponding calibrator star used as PSF reference (see text for details). 
             The best image for each source, either B12.4 ($12.47 \pm0.5\,\um$) 
or Q1 ($17.65 \pm0.44\,\um$), was selected as explained in the text.
             The colour scaling is linear in all images with black corresponding to the background level of the image and white to the brightest pixel. 
             All images where slightly smoothed with a Gaussian kernel with $\sigma=1\,$px to increase visibility (2\,px for 3C\,321 and NGC\,5135).
             The green line in each image marks the system axis PA and has a length of 200\,pc at the source distant (except for 3C\,321 where it is 1\,kpc).
            }
   \label{fig:gal_sub}
\end{figure*}

In order to better trace the extended emission, we subtract the unresolved core emission from each nucleus as shown in Fig.~\ref{fig:gal_sub}.
%
%The latter is certainly an over-simplified representation of the MIR emission structure in those AGN, which rather consists of two components, namely an unresolved core and an elongated extended structure.
%We can see this in Fig.~\ref{fig:gal_sub}, where the central unresolved core was subtracted using the corresponding reference PSF scaled to the emission peak.
In this figure, the core is actually over-subtracted (PSF peak scaled to total peak emission) to maximise the visibility of the extended emission. 
Here, and in the following analysis, we prefer the B12.4 images over Q1 ones (whenever both are available), owing to better angular resolution and sensitivity.
After the core subtraction, all objects, except NGC\,5135, show clearly extended emission.
Even in NGC\,5135 extended emission might be present as indicated by the low significance structures in the image ($\sim 3\sigma$).
Interestingly, the nuclear region of NGC\,5135 remained undetected also in recent high resolution ALMA observations tracing cold dust and molecular gas \citep{cao_alma_2018}, implying that the AGN in this object might have used up most of its fuel.
In that case, the bright \oiv emission might be coming predominantly from the kiloparsec scales, i.e, the outer NLR, as a relict of its more active past.

We verify the robustness of the extended emission detection by using the observatory provided standard stars observed in the same nights instead of our own PSF calibrators.
Owing to the PSF not being very stable in ground-based MIR observations and the observations of those standard stars being further away in time and/or sky direction, they are less suitable as PSF references.
Nevertheless, we obtain similar extended structures after PSF subtraction in all cases but NGC\,7582.
In the latter case, the PSF of observatory standard is clearly more elongated than our own and thus its subtraction leads to significant negative residuals hiding any real structure.
Therefore, we conclude from this check that the detection of extended emission is indeed robust. 

\subsection{Alignment of extended MIR emission}
To verify if the extended MIR emission is consistent with coming from the polar region of the AGN,
% we first estimate the MIR PA from the major axis of the $67\%$-flux containing ellipses from the previous section.
%The resulting values values are listed in Table~\ref{tab_sam}.
we compare the MIR PA estimated from the Gaussian fits to the system axis (SA) PAs collected from the literature.
The corresponding MIR PAs and SA PAs are listed in Table~\ref{tab_pa}.
\begin{table}
\caption{Position angles}
\label{tab_pa}
\centering
\begin{tabular}{l c c c c c}
\hline\hline
 & SA &  & host &  & MIR\\
Object & PA & Ref. & PA & Ref. & PA\\
 & [deg] &  & [deg] &  & [deg]\\
(1) & (2) & (3) & (4) & (5) & (6)\\
\hline
3C\,321 & 138 & 1, 2 & 186 & 3 & 99\\
IC\,4518W & 13 & 4 & -45 & 3 & 17\\
Mrk\,573 & 130 & 5, 6, 7, 8 & 180 & 9 & 169\\
NGC\,1365 & 125 & 10, 11, 12, 13, 14 & 20 & 3 & 102\\
NGC\,2110 & -9 & 15, 16, 17, 18 & -5 & 3 & 17\\
NGC\,5135 & 15 & 19, 20 & -60 & 9 & 1\\
NGC\,5506 & -13 & 21, 22 & 89 & 3 & 24\\
NGC\,5643 & 86 & 23, 24, 25 & 90 & 9 & 67\\
NGC\,7582 & 70 & 24, 12 & 156 & 3 & 82\\
\hline	
\end{tabular}
		                              
\begin{minipage}{1.0\columnwidth}
%\small
%\scriptsize
\normalsize
{\it -- Notes:} 
(1) object name;
(2) and (3) system axis position angle and corresponding references based on the NLR major axis from \oiii images, nuclear radio morphology, polarised broad lines, or resolved maser emission (adopted from A16).
(4) and (5) host inner structure position angle and corresponding reference (adopted from A16);
(6) mean nuclear MIR PA as determined in this work;
list of references:
1: \cite{baum_extended_1988};
2: \cite{young_scattered_1996};
3: Hyperleda;
4: \cite{rodriguez-zaurin_vlt-vimos_2011};
5: \cite{pogge_imaging_1995};
6: \cite{schmitt_comparison_1996};
7: \cite{ulvestad_radio_1984-1};
8: \cite{nagao_detection_2004};
9: this work;
10: \cite{phillips_remarkable_1983};
11: \cite{jorsater_kinematics_1984};
12: \cite{storchi-bergmann_detection_1991};
13: \cite{kristen_imaging_1997};
14: \cite{sandqvist_central_1995};
15: \cite{mulchaey_hubble_1994};
16: \cite{ulvestad_nuclear_1983};
17: \cite{nagar_radio_1999};
18: \cite{moran_transient_2007};
19: \cite{gonzalez_delgado_ultraviolet-optical_1998};
20: \cite{ulvestad_radio_1989};
21: \cite{colbert_large-scale_1996};
22: \cite{lumsden_spectropolarimetry_2004};
23: \cite{simpson_one-sided_1997};
24: \cite{morris_velocity_1985};
25: \cite{ leipski_radio_2006};

\end{minipage}
\end{table}
The latter are based on PAs measured from \oiii cones, maser disks, outflows and/or polarized broad-line emission and were collected in A16 already.
We refer the reader to that work for more details on this.

Fig.~\ref{fig:gal_sub} shows the SA PAs over-plotted on top of the MIR emission. 
Indeed, both seem to align reasonably well.
Quantitatively, the MIR PA and SA PA agree to within $23\degree$ median ($12\degree$ standard deviation) with the largest difference of $\sim39\degree$ for Mrk\,573.
Note that even for an apparent misalignment of $\sim 45\degree$, the MIR emission can still be consistent with a polar origin as shown for the Circinus AGN, where the angular difference between SA and MIR seems to be caused by the accretion disk being tilted towards one side of the polar cone \citep{stalevski_dissecting_2017}.

At the same time, the MIR emission does not align with host structure PAs (taken as well from A16 and listed in Table~\ref{tab_pa}). 
Here, the median angular difference is $62\degree$ with a standard deviation of $26\degree$.
Clearly, the preferred origin of the extended MIR emission is the polar region of AGN rather than host-related structures, verifying the findings of A16 (see also \citealt{fischer_determining_2013}).

\begin{figure}
%    \centering
%    \sidecaption
   \includegraphics[angle=0,width=0.5\columnwidth]{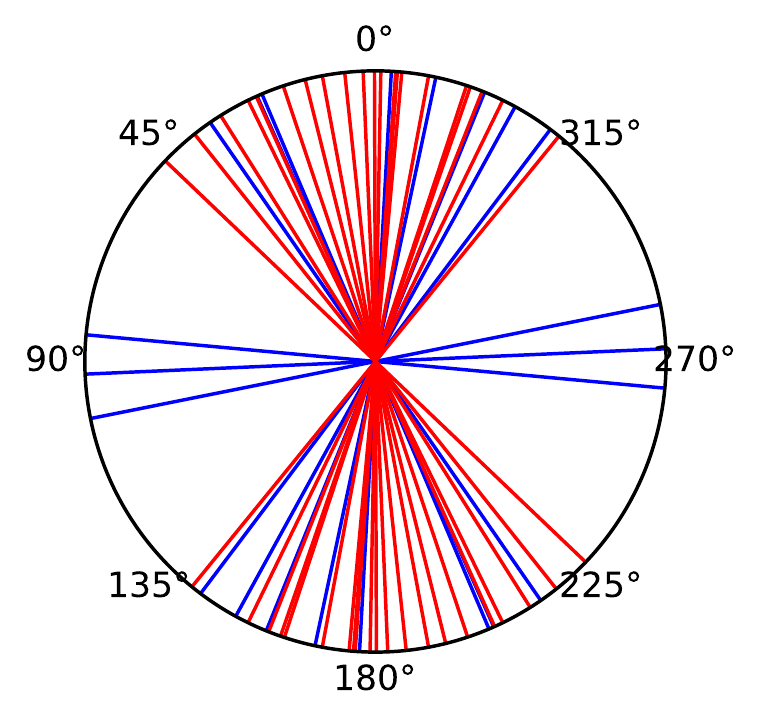}
   \includegraphics[angle=0,width=0.5\columnwidth]{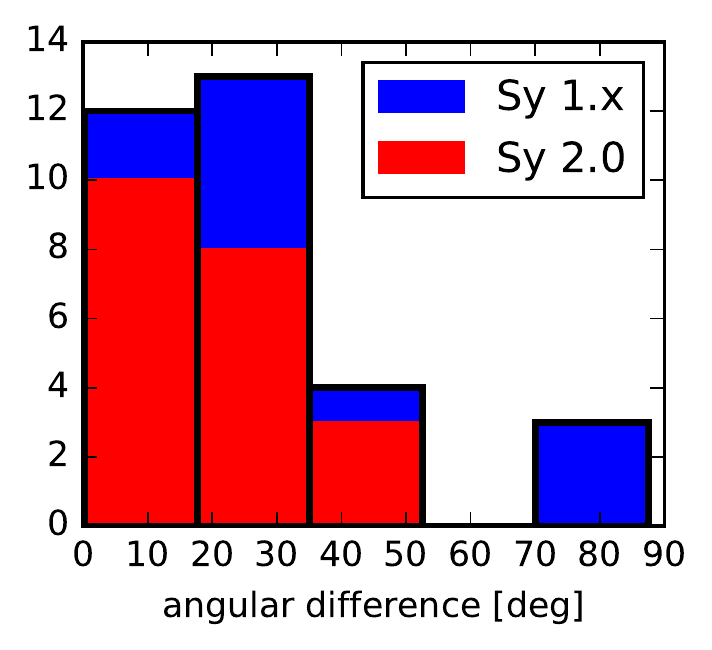}   
    \caption{
            Distribution of the angular difference between the system axis and MIR PAs for all MIR-extended Seyferts from this work and A16 combined.
             Left: angular plot showing individual objects as lines with Seyfert 1.x (2) in blue (red).
%             The thick black line shows the mean of all extended Seyferts with the black arcs indicating the standard deviation. 
%             Grey dashed lines mark the angular difference between host and MIR PAs for the same objects.
             Right: Additive histogram of absolute difference with the contribution of the Seyfert 1.x (2) objects is marked in blue (red).
 %            The distribution of the angular difference between the host and MIR PAs for all MIR-extended Seyferts is shown as grey bars.
            }
   \label{fig:pa_distr}
\end{figure}

Finally, we combine the here presented sample with the objects from A16 and show the updated distribution of MIR--SA PA difference in Fig.~\ref{fig:pa_distr}.
Adding the new sources does not change the median alignment between the MIR and SA PAs ($19\degree$), while the standard deviation slightly decreases from $27\degree$ to $23\degree$.

The above results leave little doubt on a polar origin for the extended MIR emission, while its high detection rate among our test sample verifies our hypothesis that polar dust emission on tens to hundreds of parsec scale might be an ubiquitous feature of the AGN structure. 
One has to keep in mind though that this hypothesis could only be tested for AGN in the Seyfert regime with a bright NLR, owing to the \oiv flux selection.
Given the likely origin of the polar dust as a wind/outflow, it is entirely possible, and probably even to be expected, that the presence and prominence of it will depend on the AGN fundamental parameters like the bolometric luminosity and Eddington ratio.
In order to do that (see Sect.~\ref{sec:dep}) we first need to constrain the strength and extent of the polar dust emission.

\subsection{Dominance of polar dust emission}
To understand how relevant the polar dust emission is for the energy budget of the AGN, we derive lower limits on its relative contribution, $\polagn$, towards the total MIR emission of the AGN, which we define as all MIR emission that is caused by AGN heating.
In practise, we estimate the latter by integrating all flux within a $2\arcsec$ diameter aperture in our VISIR images.
This value was chosen as compromise to encompasses as much extended emission as possible but not too much noise.
To separate the extended and unresolved emission the accurate PSF references are particularly useful because they allow us to perform a scaled point source subtraction of the unresolved core.
Here we assume the underlying brightness distribution of the extended emission to be flat and thus aim for corresponding point source scaling to obtain a smooth, centrally flat residual (see Sect.~\ref{sec:rob} for a discussion of this assumption).
Then, the extended emission is estimated as the sum of the residual emission within the $2\arcsec$ aperture, and its relative contribution,  $\extnuc$, is the ratio of the extended over the total AGN emission.

\begin{figure}
%    \centering
%    \sidecaption
   \includegraphics[angle=0,width=1.0\columnwidth]{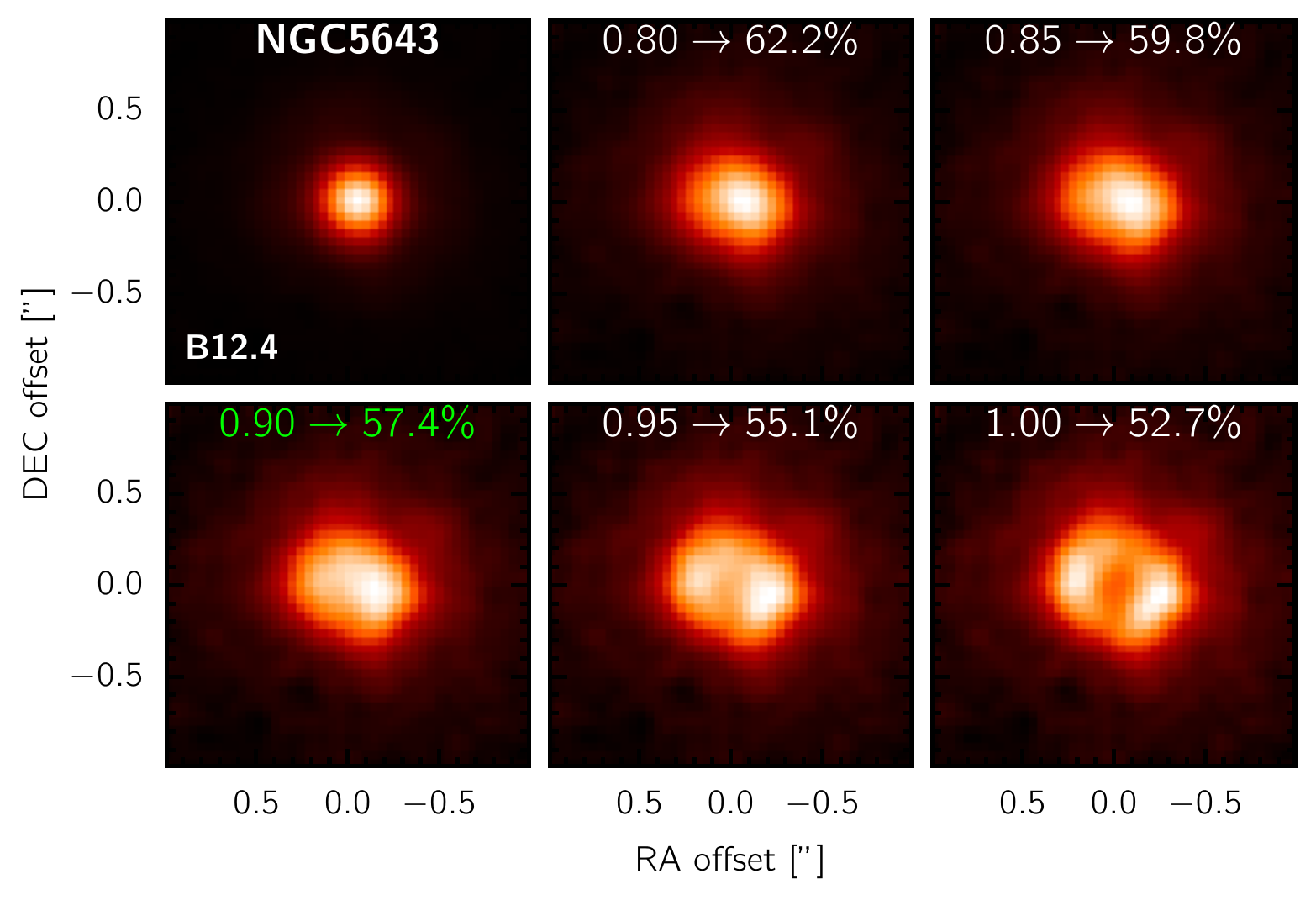}   
    \caption{
             Estimation of extended flux contribution to total AGN flux, $\extnuc$, for NGC\,5643 in the B12.4 ($12.47 \pm0.5\,\um$) filter.
             Shown is the central $2\arcsec \times 2\arcsec$ region with similar colour scaling and smoothing to Fig.~\ref{fig:gal_sub}.
             The top left image shows the original, while the bottom right shows 100$\%$ subtraction.
             The others in between show a sequence of increasing relative amplitude for point source subtraction (the first number in each subplot) from left to right.
             The two numbers in each subplot give the scaling of the point source relative to the emission peak and the resulting contribution of the residual toward the total flux, $\extnuc$, as measured in an $2\arcsec$ diameter aperture. 
             The selected scaling is marked in green.
             }
   \label{fig:sub}
\end{figure}

We find the right scaling by creating a sequence of images with increasing subtraction amplitude.
This process is illustrated in Fig.~\ref{fig:sub} for one source (NGC\,5643).
The resulting values for $\extnuc$ are listed for all objects in Table~\ref{tab_sam}, while the corresponding sequence plots, can be found in the Appendix (except for 3C\,321 where the low S/N of the image does not allow for better estimate than subtracting the point source to 100\%).
%
%\onecolumn
%\begin{landscape}
\begin{table*}
\caption{Properties of full polar dust AGN sample}
\label{tab_sam}
\centering
\begin{tabular}{l c c c c c c c c c c c}
\hline\hline
Object & Sy & D & $\log \ltw$ & $\log \lxi$) & $\log \nh$ & $\log \mbh$ & Ref. & $\log \ledd$ & $\extnuc$ & $\dmir$ & $\dmir$\\
 &  & [Mpc] & [erg/s] & [erg/s] & [cm$^{-2}$] & [$\msun$] &  &  & [\%] & [$\arcsec$] & [pc]\\
(1) & (2) & (3) & (4) & (5) & (6) & (7) & (8) & (9) & (10) & (11) & (12)\\
\hline
\hline\multicolumn{12}{c}{\textbf{new (this work)}}\\
3C\,321 & 2.0 & 460.0 & 44.6 & 43.7 & 24.0 & \dots & \dots & \dots & 70 & 1.19 & 2210\\
IC\,4518W & 2.0 & 76.1 & 43.5 & 43.0 & 23.3 & 8.8 & 1 & -2.80 & 48 & 1.09 & 389\\
Mrk\,573 & 2.0 & 73.1 & 43.5 & 43.2 & $\ge$24.2 & 6.9 & 2 & -0.81 & 49 & 0.89 & 305\\
NGC\,1365 & 1.8 & 17.9 & 42.5 & 42.1 & 23.2 & 7.3 & 3 & -2.25 & 46 & 1.26 & 109\\
NGC\,2110 & 2.0 & 35.9 & 43.1 & 42.7 & 22.5 & 8.7 & 3 & -3.16 & 70 & 1.13 & 194\\
NGC\,5135 & 2.0 & 66.0 & 43.2 & 43.2 & 24.4 & 7.0 & 4 & -0.85 & 22 & 0.83 & 258\\
NGC\,5506 & 1.9 & 31.6 & 43.4 & 43.1 & 22.4 & 7.6 & 3 & -1.59 & 32 & 0.98 & 148\\
NGC\,5643 & 2.0 & 20.9 & 42.5 & 42.1 & $\ge$24.3 & 7.0 & 5 & -2.00 & 57 & 1.06 & 106\\
NGC\,7582 & 1.8 & 23.0 & 42.9 & 42.4 & 23.1 & 7.7 & 6 & -2.46 & 44 & 0.71 & 79\\
\hline\multicolumn{12}{c}{\textbf{known (A16)}}\\
Cen\,A & 2.0 & 3.8 & 41.8 & 42.0 & 23.1 & 7.7 & 7 & -2.89 & 26 & 0.72 & 13\\
Circinus & 2.0 & 4.2 & 42.6 & 42.3 & 24.7 & 6.2 & 8 & -1.07 & 40 & 3.00 & 61\\
Cygnus\,A & 2.0 & 257.0 & 44.1 & 44.3 & 23.3 & 8.8 & 9 & -1.61 & 60 & 1.16 & 1299\\
ESO\,323-77 & 1.2 & 71.8 & 43.7 & 42.8 & 23.6 & 7.4 & 10 & -1.73 & 43 & 0.88 & 297\\
ESO\,428-14 & 2.0 & 28.2 & 42.4 & 42.6 & $\ge$24.3 & 6.8 & 3 & -1.25 & 68 & 1.75 & 236\\
IC\,5063 & 2.0 & 49.1 & 43.8 & 42.9 & 23.4 & 7.8 & 3 & -2.06 & 58 & 1.08 & 251\\
MCG-03-34-064 & 2.0 & 79.3 & 44.0 & 43.3 & 23.6 & 7.7 & 11 & -1.52 & 60 & 1.14 & 423\\
NGC\,1068 & 2.0 & 14.4 & 43.9 & 43.6 & $\ge$25.0 & 6.9 & 12 & -0.39 & 75 & 1.55 & 108\\
NGC\,1386 & 2.0 & 16.5 & 42.4 & 42.0 & $\ge$24.3 & 7.4 & 3 & -2.54 & 62 & 1.30 & 104\\
NGC\,2992 & 1.8 & 39.7 & 42.9 & 42.5 & 22.0 & 7.4 & 3 & -2.02 & 30 & 0.99 & 187\\
NGC\,3081 & 2.0 & 40.9 & 42.9 & 42.5 & 23.9 & 7.4 & 4 & -1.95 & 51 & 2.13 & 414\\
NGC\,3227 & 1.5 & 22.1 & 42.5 & 42.1 & 22.2 & 6.9 & 13 & -1.85 & 47 & 1.02 & 108\\
NGC\,3281 & 2.0 & 52.8 & 43.6 & 43.4 & 24.3 & 7.4 & 4 & -1.07 & 40 & 1.09 & 273\\
NGC\,4151 & 1.5 & 13.3 & 42.8 & 42.5 & 22.7 & 7.7 & 14 & -2.27 & 22 & 1.28 & 83\\
NGC\,4388 & 2.0 & 19.2 & 42.3 & 42.3 & 23.5 & 6.9 & 15 & -1.76 & 46 & 1.02 & 95\\
NGC\,4593 & 1.0 & 45.6 & 43.1 & 42.9 & 20.4 & 7.0 & 16 & -1.24 & 34 & 0.98 & 212\\
NGC\,5728 & 2.0 & 45.4 & 42.5 & 42.8 & 24.1 & 7.9 & 17 & -2.17 & 50 & 0.88 & 190\\
NGC\,7172 & 2.0 & 34.8 & 42.8 & 42.8 & 22.9 & 7.3 & 4 & -1.62 & 50 & 1.19 & 198\\
NGC\,7314 & 2.0 & 18.3 & 41.8 & 42.0 & 22.0 & 5.5 & 11 & -0.64 & 54 & 0.85 & 75\\
NGC\,7469 & 1.5 & 67.9 & 43.8 & 43.2 & $\le$20.7 & 7.1 & 18 & -1.00 & 30 & 0.82 & 262\\
NGC\,7674 & 2.0 & 126.0 & 44.3 & 44.0 & $\ge$24.4 & 7.3 & 3 & -0.35 & 53 & 1.35 & 779\\
\hline	
\end{tabular}
		                              
%\begin{minipage}{22cm}
\begin{minipage}{1.0\textwidth}
%\small
%\scriptsize
\normalsize
{\it -- Notes:} 
(1), (2), (3), and (4) object name, optical class, distance (D), and observed nuclear $12\um$ continuum luminosity, $\ltw$, from \cite{asmus_subarcsecond_2014};
(5) and (6) intrinsic 2-10\,keV X-ray luminosity, $\lxi$, and obscuring X-ray column density, $\nh$, from \cite{asmus_subarcsecond_2015}, except NGC\,5643 which was taken from \cite{annuar_nustar_2015};
(7), (8) and (9) estimated black hole mass, $\mbh$, corresponding reference and resulting Eddington ratios, $\ledd$, assuming a bolometric luminosity $\lbol = 10 \lxi$;
(10) resolved fraction of the total nuclear MIR emission as estimated in this work from PSF subtraction;
(11) and (12) measured major axis diameters containing 80\% of the total AGN MIR emission in angular and absolute units, $\dmira$ and $\dmirp$, respectively;
list of references:
1: \cite{koss_bat_2017};
2: \cite{bian_eddington_2007};
3: Hyperleda;
4: \cite{garcia-rissmann_atlas_2005};
5: \cite{beifiori_upper_2009};
6: \cite{wold_nuclear_2006};
7: \cite{cappellari_mass_2009};
8: \cite{greenhill_warped_2003};
9: \cite{gebhardt_axisymmetric_2003};
10: \cite{wang_unified_2007};
11: \cite{cid_fernandes_star_2004};
12: \cite{lodato_non-keplerian_2003};
13: \cite{denney_reverberation_2010};
14: \cite{winter_x-ray_2009};
15: \cite{kuo_megamaser_2011};
16: \cite{denney_mass_2006};
17: \cite{mcelroy_catalog_1995};
18: \cite{peterson_central_2004};

\end{minipage}
\end{table*}
%\end{landscape}
%
On median, $\extnuc$ is 48\% (standard deviation 18\%), ranging from 22\%, in NGC\,5135, to 70\%, in NGC\,2110.

\subsubsection{Robustness of method}\label{sec:rob}
This method, despite being somewhat arbitrary and subjective has been commonly used in the literature to separate unresolved and extended components due to the lack of a  more objective approach \citep[e.g.,][]{radomski_high-resolution_2002, radomski_resolved_2003, soifer_high_2003, ramos_almeida_infrared_2009}.
We will address the possible caveats of this method with some tests.
Firstly, we verify that we obtain similar results for using the observatory provided standard stars.
Using those, the resolved fractions are on average 4,5\% smaller but consistent within 1$\sigma$ (standard deviation is 5\%).

Secondly, we subtract our PSF references from the observatory standards to test the amplitude of any artefacts of the subtraction method. 
The resulting $\extnuc$ measured in the same way as for our AGN, are on average only 5\% (standard deviation 11\%), so much smaller than our resolved fractions measured for the AGN.

\begin{figure}
%    \centering
%    \sidecaption
   \includegraphics[angle=0,width=1.0\columnwidth]{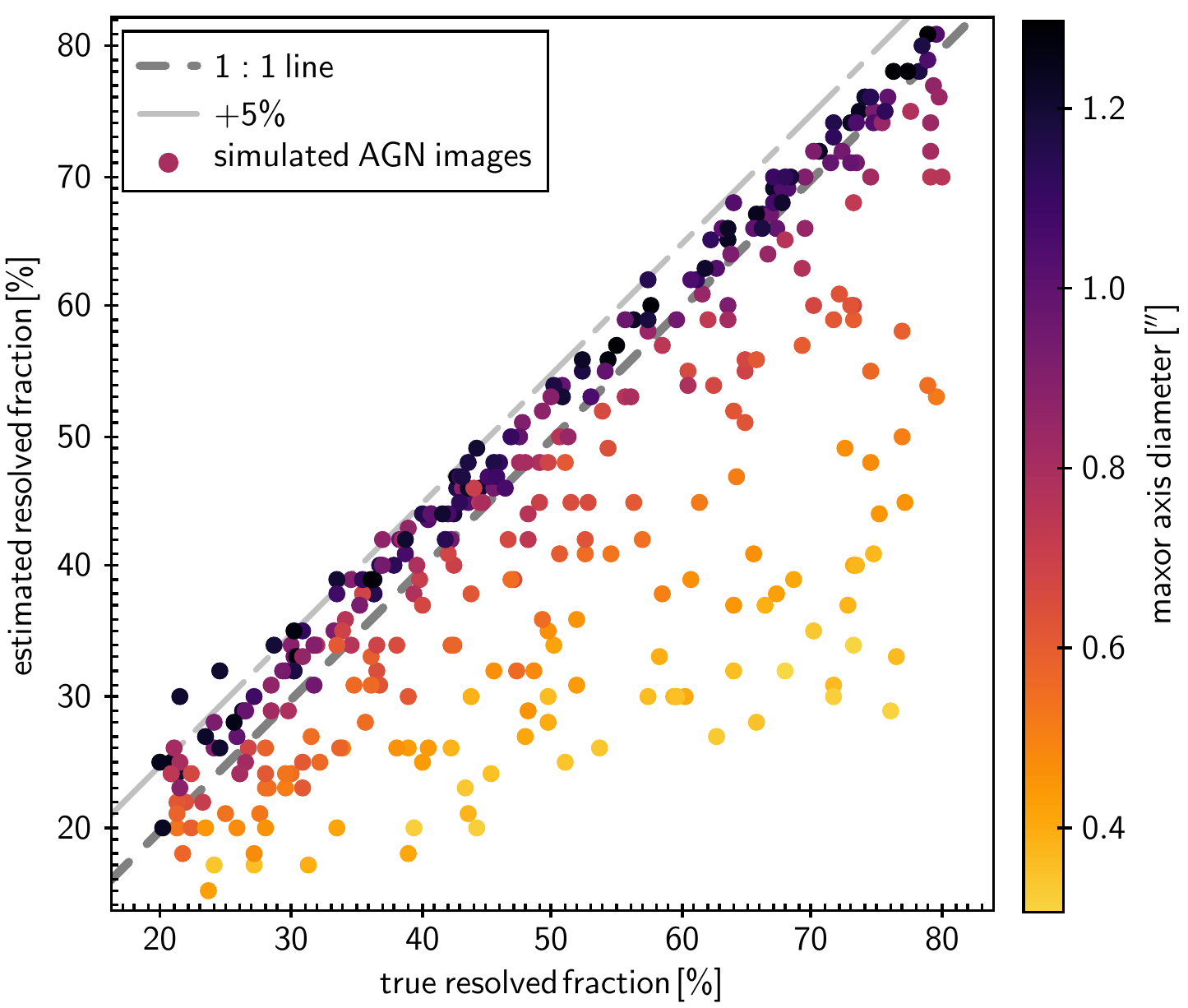}   
    \caption{
             Estimation of extended flux contribution to total flux, $\extnuc$, for 350 simulated images in the B12.4 filter.
             The random, hidden input $\extnuc$ is shown on the x axis while the estimated value is on the y axis. 
             The symbols are colour coded by the random size of the ellipsoid representing the extended emission. 
             The dot-dashed and dashed line show the 1:1 and the 1:1 $+5\%$ relation, respectively.
             For details, see Sect.~\ref{sec:rob}.
             }
   \label{fig:sim}
\end{figure}

Finally, we test whether there is a systematic bias in our flux estimates owing to the tuning ''by eye''. 
For this purpose, we create a simulation where we add a PSF-convolved elliptical extended component with random amplitude, size and orientation to the B12.4 image of NGC 5135.
We chose this image because it does not show significant extended emission and matches in terms of S/N to our other AGN images better than any calibrator star observation. 
Furthermore, the circumnuclear emission in NGC 5135 presents are harder challenge to test our method.
The range for the amplitude and size of the ellipsoid are matched to the detected extended emission in our AGN but the exact generated values remain hidden during the test.
We then measure the extended flux contribution, $\extnuc$, in the same way as for our real observations and calculate the difference of real value minus estimated value. We repeat this 350 times, at which stage the resulting statistics have converged to $<1\%$.
The resulting distribution of the real and estimated $\extnuc$ is shown in Fig.~\ref{fig:sim}.
On average, our method underestimates the true $\extnuc$ by 6\% with a standard deviation of 11\%. The distribution of difference between true and estimated value is however highly asymmetric: the with 80\% of the values lying between -3\% and +23\%. 
This asymmetry is caused by the fact that we tend to underestimate the contribution of the extended component if it is compact as shown by the colour coding in Fig.~\ref{fig:sim}.
The compacter the extended emission, the more of it is accounted to the point source by our method mistake.
On the other hand, the method only rarely overestimates the contribution of the extended component by more than 5\% ($<1$\% of the cases). 

Finally, one could question the assumption of the extended emission to have a smooth, flat profile in the central $\sim 0.4\arcsec$ region.
Firstly, even in the worst case scenario, namely that it actually would have zero emission in the centre, which is the same as the 100\% subtraction shown in Fig.~\ref{fig:gal_sub}, we would get only $7\%$ smaller $\extnuc$ on average, thus not changing our conclusions.
Secondly, the size of the VISIR PSF would require the underlying brightness distribution to have a "hole" with a projected radius of $>0.2$\arcsec\, ($>40$\,pc at the median object distance) to produce a central drop in the PSF-convolved brightness distribution in the observed images.
However, such a scenario seems unlikely if the extended dust is part of the AGN structure and is centrally heated, as the previous results indicate.
Therefore, we believe our flat profile assumption to be reasonable.

In summary, we conclude that our method is robust to within $10\%$ on average and, if anything, rather underestimates the extended flux for our objects.

\subsubsection{Application to full sample}
\begin{figure}
    \centering
%    \sidecaption
   \includegraphics[angle=0,width=0.6\columnwidth]{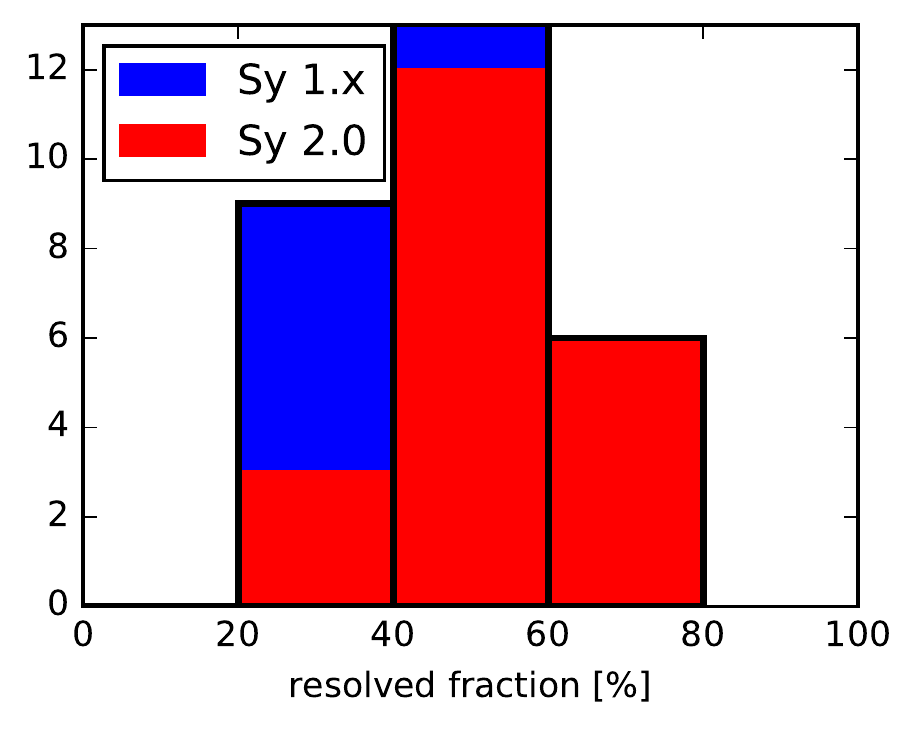}   
    \caption{
            Distribution of minimum relative contribution of the polar extended emission, $\extnuc$, for all MIR-extended Seyferts from this work and A16 combined.
The histogram is additive with the contribution of the Seyfert 1.x (2) objects marked in blue (red).
 %            The distribution of the angular difference between the host and MIR PAs for all MIR-extended Seyferts is shown as grey bars.
            }
   \label{fig:ext_distr}
\end{figure}

We also employ the same method to similar imaging data of the A16 sample to obtain improved values of $\extnuc$ for all AGN with detected polar dust emission.
Details on those data will be presented in a future work  (based on ESO programmes 099.B-0235, 100.B-0056, and 101.B-0334; PI H\"onig; Asmus et al., in prep.).
Furthermore, we add two objects, IC\,4329A and NGC\,4151, which were discussed in A16 to have good evidence for polar dust emission as well (e.g.,  \citealt{radomski_resolved_2003}). 
The resulting distribution of $\extnuc$ is shown in Fig.~\ref{fig:ext_distr}.
The median value of $\extnuc$ is $49\%$ with a standard deviation of $14\%$, whereas Seyfert~2 AGN have on average a larger $\extnuc$ than Seyfert~1.x AGN ($53\%$ versus $34\%$).
This trend is explained by Seyfert~2s having larger inclinations with respect to our line of sight, leading to larger projected angular sizes of the polar structures.

\begin{figure}
%    \centering
%    \sidecaption
   \includegraphics[angle=0,width=1.0\columnwidth]{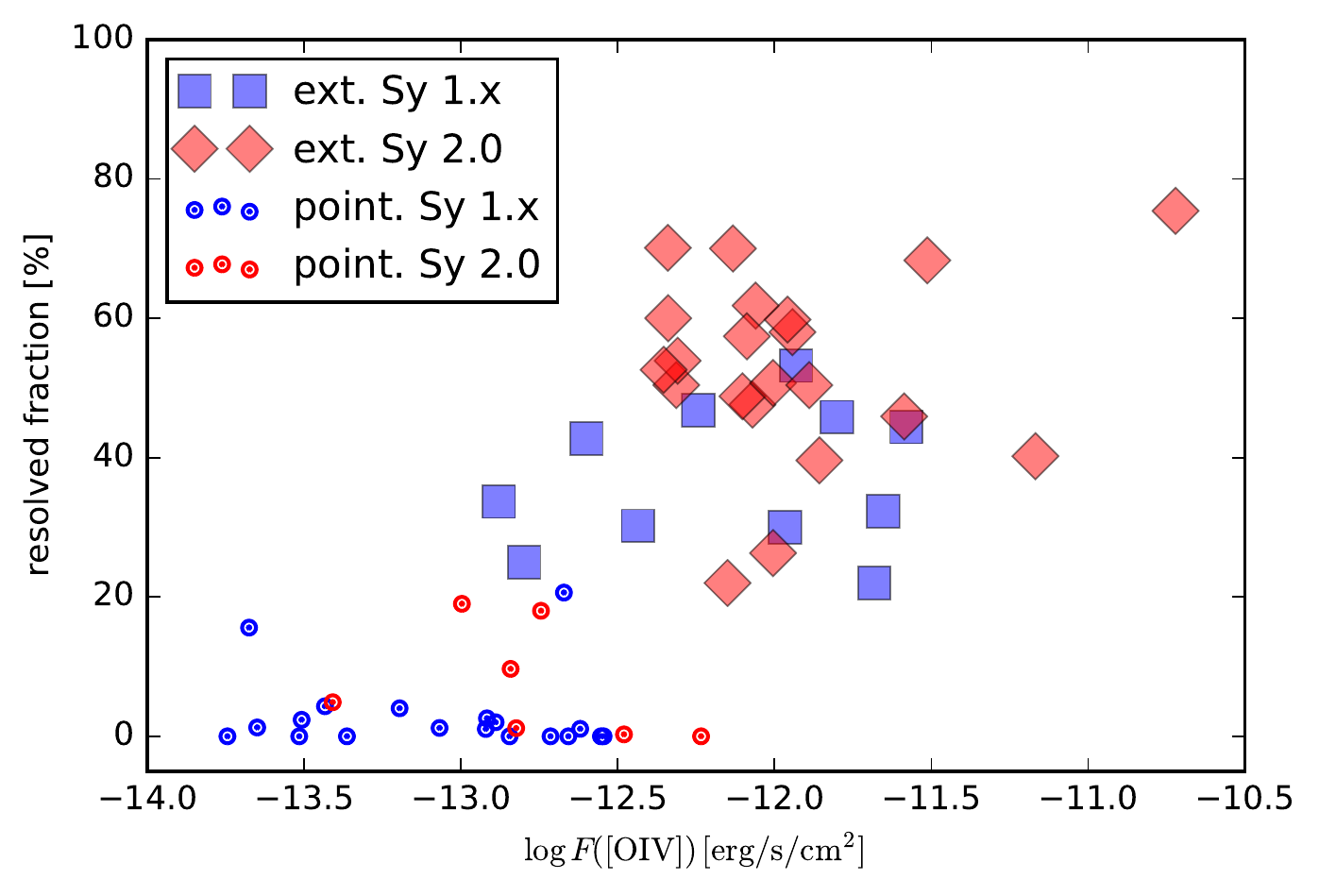}   
    \caption{
    Minimum relative contribution of the polar extended emission, $\extnuc$, over the \oiv flux, $\foiv$, for all MIR-extended Seyferts from this work and A16 combined.
            Extended Seyfert~2 objects are marked by red diamonds while extended Seyfert~1.x are blue squares. 
           For comparison, the point-like Seyferts from the total sample of A16 are shown with $\odot$ symbols (blue: Sy~1.x; red: Sy~2).
            }
   \label{fig:oiv}
\end{figure}

We also update the found trend of increasing resolved fraction with increasing \oiv flux, which was used for selecting the sample (Fig.~\ref{fig:oiv}).
The significance of this trend increases slightly with adding the new sources, i.e., the null hypothesis probability of the corresponding Spearman rank test drops from  $3\cdot 10^{-7}$ to $6\cdot 10^{-9}$ while the rank increases slightly from 0.66 to 0.68.

We emphasize that $\extnuc$ only provides a lower limit for the actual contribution of the polar dust emission, $\polagn$, because a large part of the latter remains unresolved with VISIR.
This is obvious from the VLTI/MIDI results that show prominent polar extended MIR emission at an order of magnitude higher angular resolution  in all objects with sufficient u,v coverage \citep{lopez-gonzaga_mid-infrared_2016}.
In general, the polar dust emission turns out to be the dominating component on these scales with an median flux contribution of $\polmid = 67 \pm13$\%.
Now, the VISIR unresolved component corresponds to the total flux as seen MIDI, i.e., the value at baseline length 0 (see e.g., \citealt{burtscher_diversity_2013} for details).
Therefore, the actual contribution of the polar dust emission is given by the combination of the resolved fraction in VISIR and the polar component contribution in MIDI, i.e., $\polagn/[\%] = \extnuc + (100-\extnuc)(\polmid/100) $, which is $83\%$ for the median values above.
This implies the polar dust on average dominates the total MIR emission of AGN.

Here, we do not explicitly take into account obscuration effects in the mid-infrared.
For example, \cite{stalevski_dust_2016} find that self-shielding of clumpy dust distributions around AGN can cause a factor $\sim 2$ decrease of the observed 12$\,\um$ emission for close to edge-on sight lines.
However, such hiding of obscuring dust emission would occur only on parsec-scales, so affect only MIDI measurements.
Therefore, even if the parsec-scale component would be intrinsically brighter by a factor of two, the median of  $\polmid$ would become $51\%$ for the values from \cite{lopez-gonzaga_mid-infrared_2016}, i.e., the MIR emission would still be dominated by the polar dust.
%This would only change if $\polmid<2\%$, i.e., the parsec-scale component  be intrinsically brighter by a factor $>140$.

\subsection{Extent of polar dust emission}
In this section, we investigate the extent of the polar dust structures.
Despite the relatively long exposure times of the new VISIR images presented here (30\,min on source), the detection of the extended MIR structures is still limited by low signal-to-noise ratios.
This is because ground-based MIR instruments have a relatively poor sensitivity for extended emission owing to the background noise increasing with the diameter of the telescope at the same rate as the collected photons from the source.
Thus, the full extent of the polar dust emission can only be probed from space, i.e., with the {\it James Webb Space Telescope} (\jwst; \citealt{gardner_james_2006}).
However, we can at least derive lower limits on the former with the VISIR images.
For this purpose, we measure the extent of the region which contains 80\% of the total nuclear emission by masking the rest and fitting an ellipse.
This threshold was chosen in analogy to the FWHM of a Gaussian and should roughly account for the PSF-widening under the assumption that the surface brightness of the polar dust drops smoothly.
We use the major axis diameter of the fitted ellipse,  $\dmir$, as measure for the MIR size.
Note that this method gives larger sizes than the major axis FWHM values of the previous Gauss fits because the latter is usually dominated by the unresolved core and, thus, misses a significant part of the extended emission.

\begin{figure}
%    \centering
%    \sidecaption
   \includegraphics[angle=0,width=0.5\columnwidth]{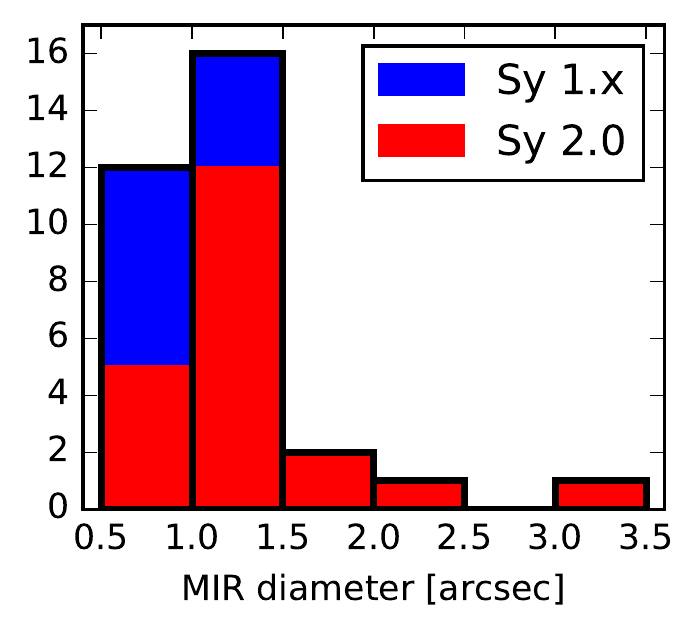}
   \includegraphics[angle=0,width=0.5\columnwidth]{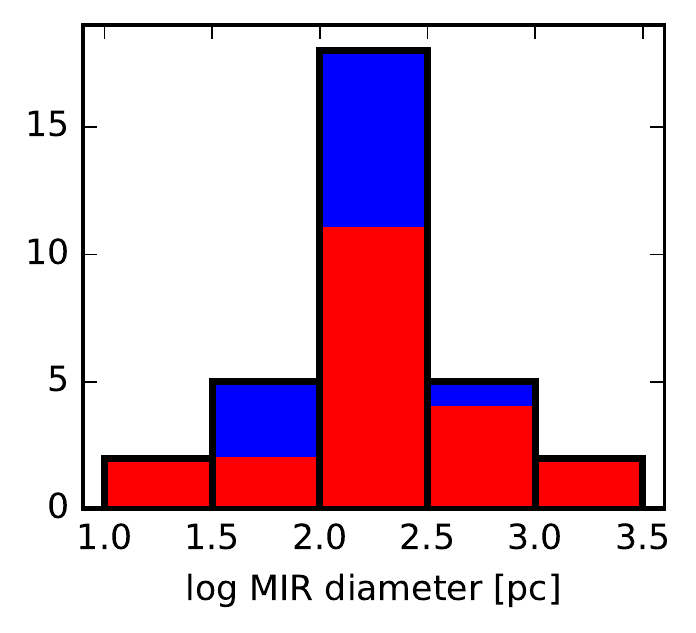}   
    \caption{
            Distribution of the MIR major axis diameters containing 80\% of the total nuclear flux, $\dmir$, for all MIR-extended Seyferts from this work and A16 combined.
            The contributions of the Seyfert 1.x (2) are marked in blue (red).
             Left: sizes are in angular units.
             Right: size are in absolute units.
            }
   \label{fig:siz_distr}
\end{figure}

The distribution of $\dmir$ for all polar extended AGN is shown in Fig.~\ref{fig:siz_distr}, while the individual values are listed in Table~\ref{tab_sam}.
As explained above, the relatively poor sensitivity for extended emission leads to the resulting angular sizes falling into a narrow range for most objects, namely $\dmir \sim 1.1\arcsec \pm 0.3\arcsec$.
Converted into absolute units at the object distance, $D$, $\dmir$ ranges from 13\,pc in Cen\,A to 2.2\,kpc in 3C\,321 with a median of 198\,pc (standard deviation 418\,pc; right side of Fig.~\ref{fig:siz_distr}).
%It is not a coincidence that Cen\,A is the closest object and 3C\,321 the most distant.
%Namely, the measured sizes correlate strongly with the object distance despite many of the objects apparently being well-resolved (sizes $>2\times$ PSF FWHM).
While the absolute sizes are more physical meaningful for the investigation of any trends of the polar MIR emission size with other AGN parameters, we need to keep in mind that the $\dmirp$ are still lower limits that, owing to the above caveat, are dominated by the object distance, i.e., $\dmirp$ and $D$ are strongly correlated.

%This linear correlation is stronger than with any other quantity and thus suggests that the measured MIR sizes are dominated by our fixed angular resolution. 
%A PSF deconvolution would be required to correct for this but the S/N of our images is too low in most cases for reliable results.
%Therefore, we will need to correct for the distance effect if we want to

% - try with PSF deconvolution --> does not work (S/N too low?)

\subsection{Dependence on AGN-fundamental parameters}\label{sec:dep}
As already mentioned, the most likely explanation for significant amount of dust in the polar region of the AGN is a roughly vertically launched dusty wind driven by radiation from the accretion disk.
In this scenario, it is naturally expected that the amount of dust and the size of the outflow will be connected with the AGN-fundamental parameters of luminosity and/or accretion rate.
Therefore, we test for any empirical correlations of our derived lower limits for the contribution and size of the polar dust emission, $\extnuc$ and $\dmir$, with the bolometric luminosity, $\lbol$ and the Eddington ratio, $\ledd$.
Here,  $\lbol$ is approximated simply as $\lbol = 10 \lxi$, following, e.g.,  \cite{vasudevan_piecing_2007} and \cite{vasudevan_optical--x-ray_2009} with $\lxi$ the intrinsic 2-10\,keV X-ray luminosity, while $\ledd = \lbol / \Ledd$ with $\ledd = 1.26 \cdot 10^{38} (\mbh /\msun) $\,erg/s the Eddington luminosity and $\mbh$ the black hole mass.
The values for $\lxi$ and $\mbh$ are collected from the literature and are listed in Table~\ref{tab_sam}.

We tested for correlations between all the above quantities, expressing $\dmir$ in angular, absolute and relative units.
For the latter, we express $\dmir$ in units of the sublimation radius, $\rsub$, which is approximated as $\rsub = 0.47 \sqrt{ L_\mathrm{bol} / 10^{46}\mathrm{erg}\,\mathrm{s}^{-1}}$\,pc, under the assumption that the innermost dust is dominated by graphite grains of $0.04\um$ size with a sublimation temperature of 1500\,K \citep{kishimoto_innermost_2007}.

Unfortunately, our sample spans only a narrow range in luminosity for a given distance, leading to artificially strong correlations whenever distance and luminosity are involved.
Once this caveat is taken into account, no statistically significant correlations remain. 
This does not mean that there are no intrinsic correlations between, e.g., $\lbol$ and $\extnuc$ or $\dmir$ but they remain hidden by the dominating distance effects in our sample.

\begin{figure}
%    \centering
%    \sidecaption
   \includegraphics[angle=0,width=1.0\columnwidth]{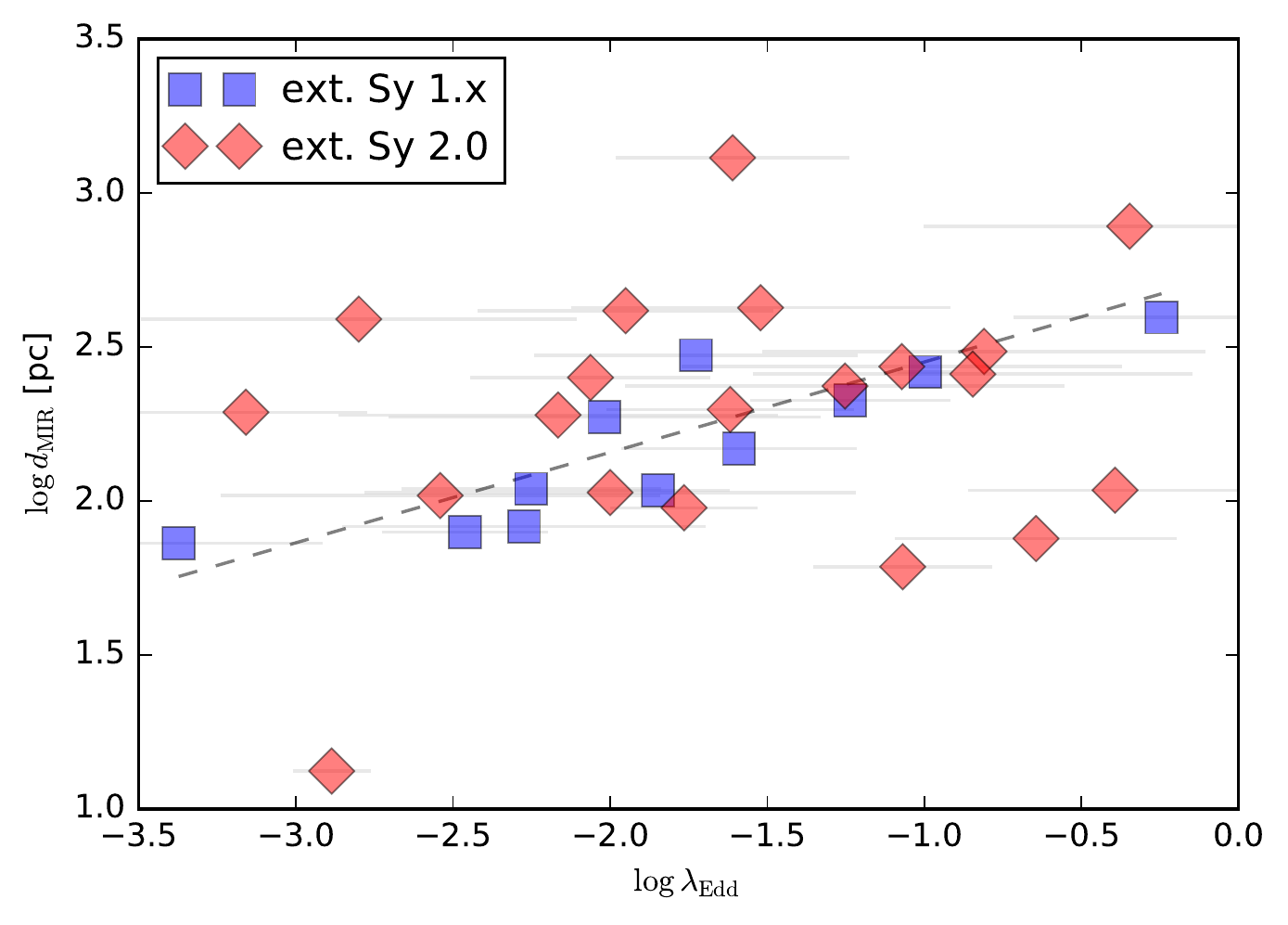}   
    \caption{
    MIR diameter, $\dmirp$ versus Eddington ratio, $\ledd$, in logarithmic space for all MIR-extended Seyferts from this work and A16 combined.
            Extended Seyfert~2 objects are marked by red diamonds while extended Seyfert~1.x are blue squares. 
            The linear fit described in the main text is shown as black dashed line.
            }
   \label{fig:sizedd}
\end{figure}

That being said, a possible correlation between $\dmirp$ and the Eddington ratio is found (Fig.~\ref{fig:sizedd}). 
It can be represented by a linear relation in logarithmic space with a slope of $\sim 0.29 \pm 0.15$ according to a fit using the ordinary least squares method with $\ledd$ as independent variable.
Here, the uncertainties are dominated by the black hole mass estimates which were collected from different methods but often with uncertainties of the order of $0.4\,$dex.
We quantify the significance of this relation using the Kendall's Tau which is $\tauK = 0.3$ with a null hypothesis probability of $\pK = 0.018$.
While this is usually not regarded as formally significant, we note the following features in support of the correlation being real.
First, the correlation properties do not change if we use the \oiv luminosity to estimate $\lbol$ instead of the X-ray luminosity.
Second, the correlation of $D$ with $\ledd$ is weaker, although just slightly ($\tauK = 0.28$ and $\pK = 0.03$).
A correlation of the Eddington ratio with the distance would also be surprising owing to the distance cancelling out in the luminosity ratio used to calculate $\ledd$.
And third and somewhat surprising, the correlation is stronger for the Seyfert~1.x only ($\tauK = 0.78$ and $\log \pK = -3.1$).

The latter finding is quite puzzling because the Seyfert 1.x sources are expected to be seen rather face-on leading to smaller and less reliable projected sizes.
As said, there is a strong correlation of $\dmirp$ with $D$ but it is in fact weaker for the Seyfert~1.x than for the Seyfert~2, which disfavours a distance effect being at work here.
Next, one might expect a correlation of the black hole mass with the distance because the estimation methods for $\mbh$ might depend on intrinsic resolution.
However, there is no such correlation present in our data, in particular for the Seyfert~1.x.
On the other hand, we note that the Eddington ratio is much more uncertain for the Seyfert~2s because both the black hole mass and bolometric luminosity estimates are less reliable, owing to being indirect and absorption correction dependent, respectively. 
This could explain the larger scatter of the Seyfert~2s, masking any correlation for them.

Certainly, the found trend of increasing absolute MIR size with Eddington ratio should be regarded with a lot of caution. 
It is unexpected and based on few objects and at the limit of what can be done with the VISIR data.
However, we failed to identify any artificial cause for such a correlation.
If it would prove to be true, then it would indicate a new relation of the polar dust structure with an AGN fundamental parameter that might provide important constraints for the modelling and thus further understanding of the polar emission and its underlying physics. 
Therefore, we discuss its implications a bit further below.

Naively, one would expect that the size of the polar MIR emitting structure scales with the luminosity.
This would then imply the Eddington ratio to correlate with $\dmirr$ rather than $\dmirp$, as found.
However, as said, our dataset does not allow us to isolate any such correlation.
At least for the less-inclined Seyfert~1.x, a possible explanation for the found correlation could be a widening of the opening angle with increasing Eddington ratio, leading to larger projected diameters.
Related to that, \cite{ezhikode_determining_2017} found a decrease in the dust covering factor, i.e., the ratio of the infrared to bolometric luminosity, with increasing Eddington ratio.
They also found that this trend is stronger with $\ledd$ than with $\lbol$. 
Similar results were obtained for the nuclear obscuration seen in X-rays \citep{fabian_effect_2008, fabian_radiation_2009,  ricci_close_2017}.
In particular in the latter work, a deficit of obscured systems at high Eddington ratios could be shown, implying that the radiation pressure of the AGN becomes so strong at high $\ledd$ that the non-equatorial regions are cleared from obscuring material. 
Both results imply an increase in the opening angle of the dusty outflow, which might also explain our finding.
On the other hand, an increase of the opening angle would imply a decrease of the MIR--X-ray emission ratio with increasing $\ledd$, which we do not see in our sample.
Alternatively, the larger MIR diameter for larger Eddington ratios might be caused by material blown to larger scales.
Detailed modelling of the dust structure in the polar dust AGN might allow us to further distinguish between the above scenarios but is beyond the scope of this work.

% - make a composite image, combining the point source subtracted (or even better PSF-deconvolved) and scaled (with the square-root of the luminosity) images. 

\section{Summary \& Conclusions}\label{sec:concl}
We have observed 9 nearby, obscured Seyfert-type AGN with deep subarcsecond resolution MIR images to look for polar dust emission, which was predicted to be present in all of them given by their \oiv brightness and inclination. 
Extended emission was detected in 8 out of 9 cases at more than $3\sigma$, and in the remaining case at $\sim2\sigma$ level.
We showed that this extended MIR emission is aligned with the system axis of the AGN in projection.
Thus, the above prediction was confirmed, with the most likely origin of the extended MIR emission being the proposed polar dusty wind.
This result suggests that polar dust might be an ubiquitous feature of the AGN structure in general, at least for the probed regime of luminosities and Eddington ratios, which extends over the full Seyfert regime ($42 \lesssim \log \lbol \lesssim 45$ and $-3 \lesssim \log \ledd \lesssim 0$).

Furthermore, the analyses of the combined samples of all  AGN with detected polar dust emission verifies that the majority of the total MIR radiation produced by the AGN is originating from these polar regions. 
Owing to the polar dust being on similar scales as the NLR, encompassing it at least partly, it has to be optically thin on average to allow us to see the NLR in inclined AGN. 
As a consequence, the MIR emission should be quite isotropic.
This would then explain, among other observations, the similarity of the MIR--X-ray emission ratio of unobscured and obscured AGN as found in the form of a tight MIR--X-ray luminosity relation (e.g., \citealt{gandhi_resolving_2009, asmus_subarcsecond_2015, mateos_revisiting_2015}).
Furthermore, it would cast doubt on parameters derived from spectral energy distribution fitting in the infrared with clumpy torus models, in particular the covering factor of optically thick material. 

In other words, the MIR is probably not tracing the main obscuring structure in AGN.
It might not even trace the bulk of the dusty material, which may reside in the much cooler structures that are shielded from the direct AGN emission and now are resolved with ALMA in several nearby AGN \citep{garcia-burillo_molecular_2014,garcia-burillo_alma_2016, gallimore_high-velocity_2016, izumi_circumnuclear_2018, alonso-herrero_resolving_2018, combes_alma_2019}.

Finally, we found a tentative trend of increasing size of the polar MIR emitting region with increasing Eddington ratio which is significant at least for Seyfert 1.x AGN. 
While we caution of an unidentified artificial origin, one possible physical explanation would be a widening of the opening angle of the dusty outflow with increasing $\ledd$.

To better understand the dust structure in AGN, detailed modelling will be required.
Here, as demonstrated in \cite{stalevski_dissecting_2017} and \cite{stalevski_dissecting_2019}, taking into account the spatial information when comparing data to models is absolutely critical to break the degeneracies that occur if one only fits the integrated spectral energy distributions.
Therefore, we plan to test both the polar dust cone and clumpy torus scenarios for the whole polar dust AGN sample in a follow-up work, where we will use the most recent radiative transfer models by \cite{honig_dusty_2017} and \cite{stalevski_dissecting_2017} to perform a combined fitting of the spectral and spatial properties of the sample (Asmus et al., in prep.).

Eventually, \jwstt will enable us significantly advance on the results presented here thanks to its vastly better sensitivity to extended emission, allowing for larger samples and more accurate size estimates at the same time.

\section*{Acknowledgements}
DA thanks the anonymous referee for critical but constructive comments that helped to a more balanced discussion.
DA as well thanks M.~Kishimoto, M.~Stalevski and K.R.W.~Tristram for very helpful and constructive comments on the manuscript, and S.F.~H\"onig, P.~Gandhi, and C.~Ricci for their support and feedback at various stages of this work.
Furthermore, DA thanks the NED team for providing the table with all the \oiv measurements included in NED.
Based on European Southern Observatory (ESO) observing programme
099.B-0044.
DA acknowledges funding through the European Union’s Horizon 2020 and Innovation programme 
under the Marie Sklodowska-Curie grant agreement no. 793499 (DUSTDEVILS).
%
%SFH acknowledges support from the Horizon 2020 ERC Starting Grant DUST-IN-THE-WIND (ERC-2015-StG-677117).
%
%
This research made use of the NASA/IPAC Extragalactic Database
(NED), which is operated by the Jet Propulsion Laboratory, California Institute of Technology, under contract with the National Aeronautics and Space Administration. 
This research made use of Astropy, a community-developed core Python package for Astronomy \citep{astropy_collaboration_astropy:_2013}.
This research has made use of NASA's Astrophysics Data System.
%
% 
% for the bibliography, at the end
%\bibliographystyle{aa} % style aa.bst
\bibliographystyle{mn2e} 
\bibliography{my_lib_ref.bib} % your references Yourfile.bib

\appendix
%\twocolumn

\begin{figure*}
%    \centering
%    \sidecaption
   \includegraphics[angle=0,width=1.0\columnwidth]{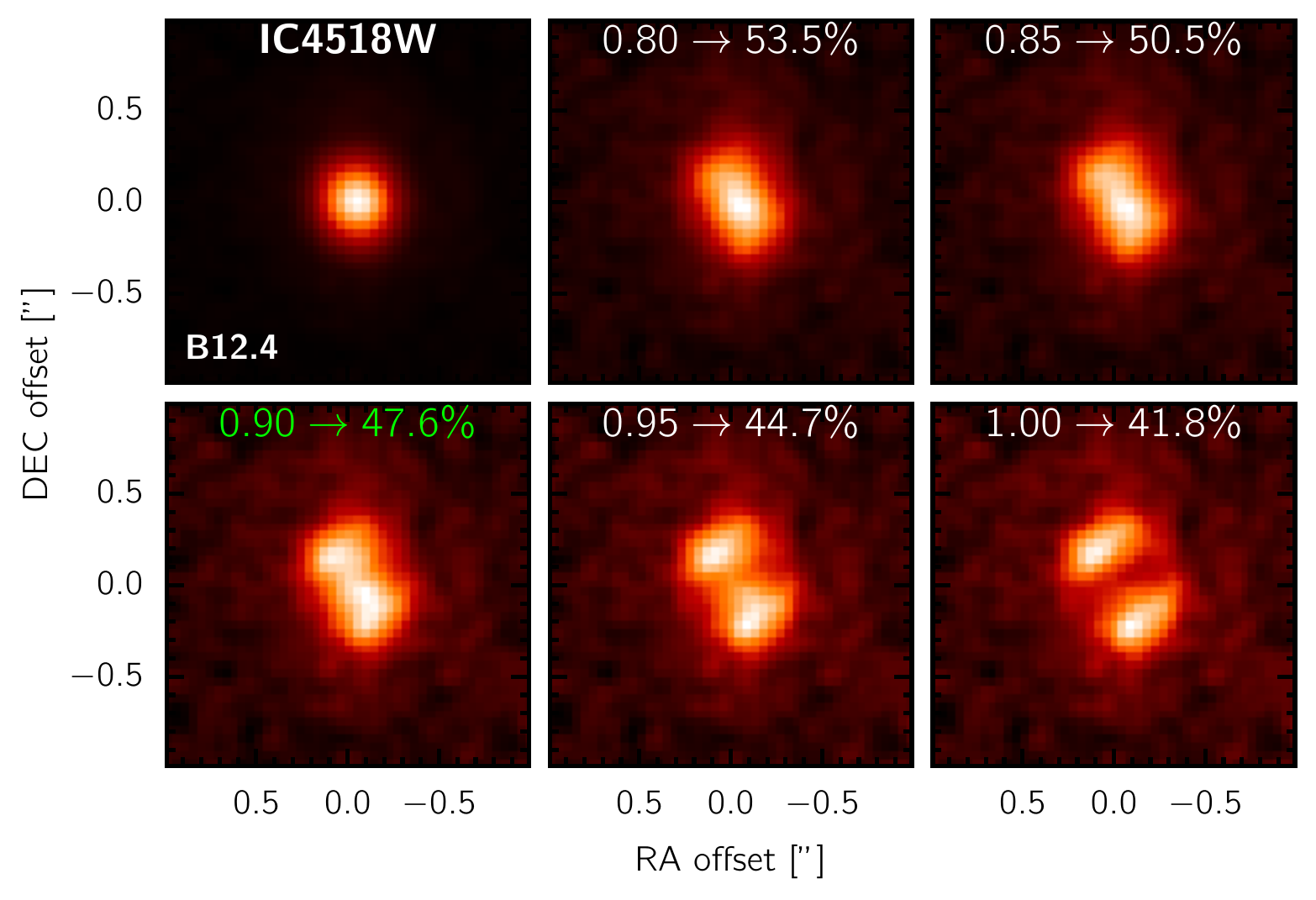}  
   \includegraphics[angle=0,width=1.0\columnwidth]{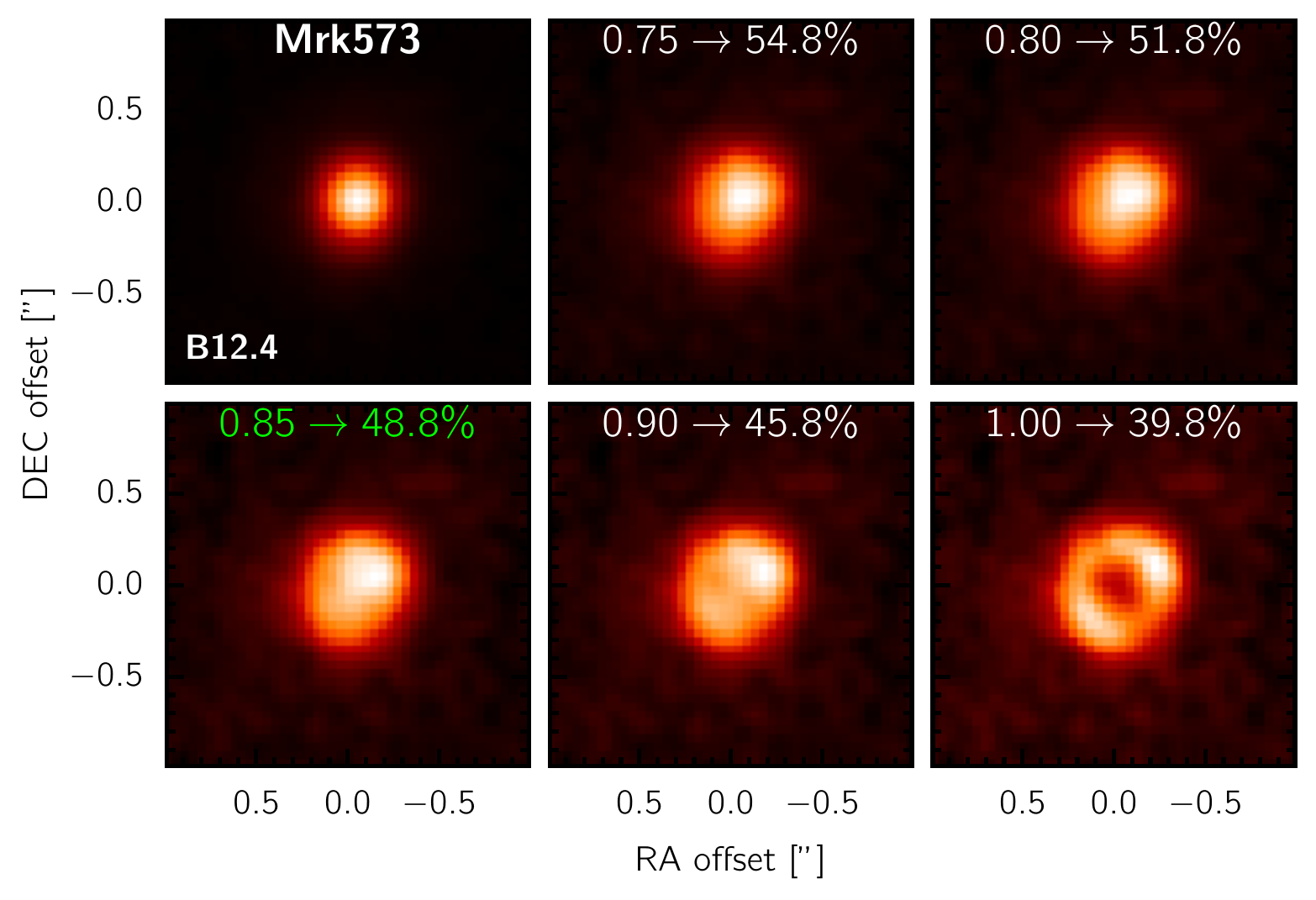}  
      \includegraphics[angle=0,width=1.0\columnwidth]{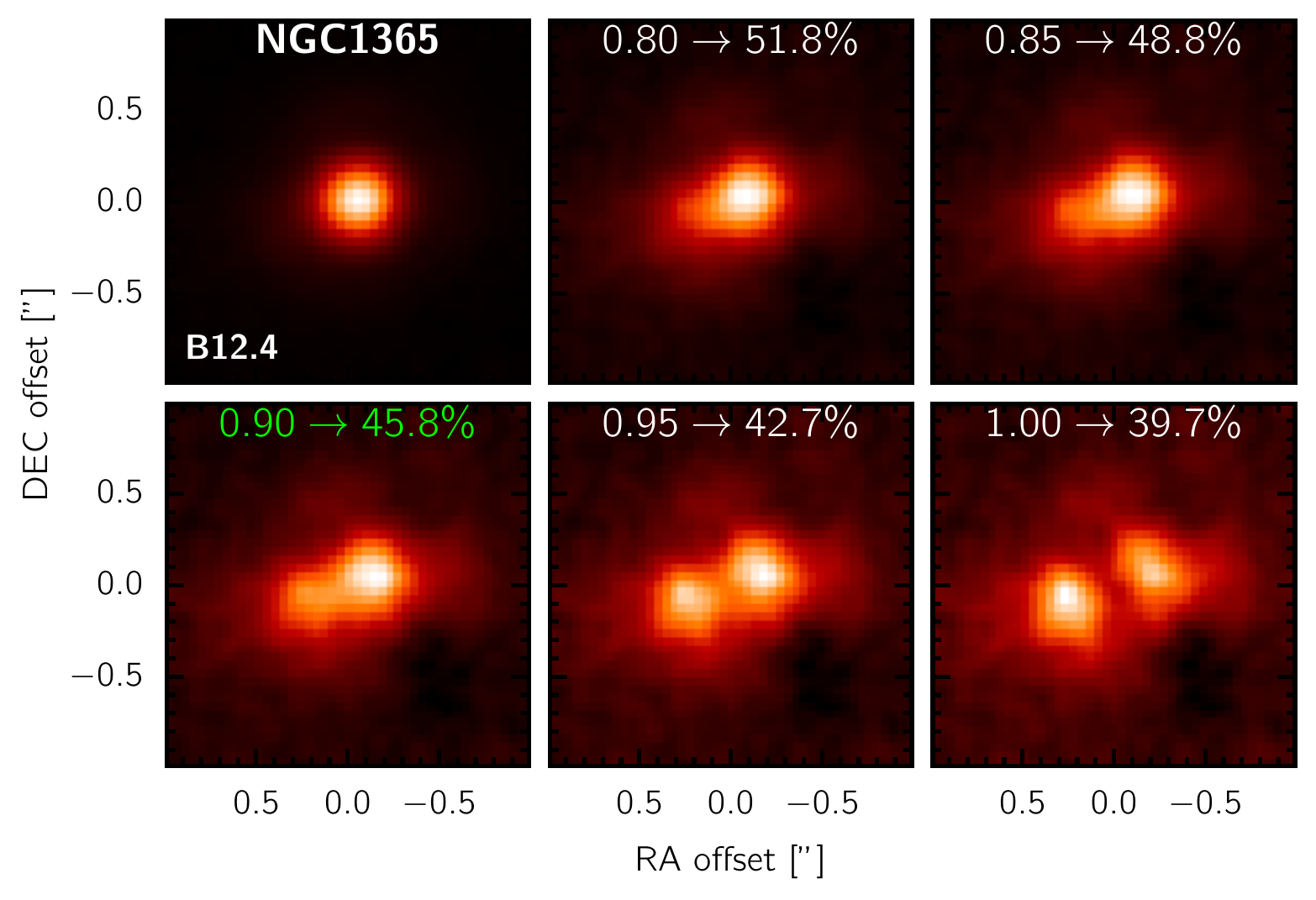}  
         \includegraphics[angle=0,width=1.0\columnwidth]{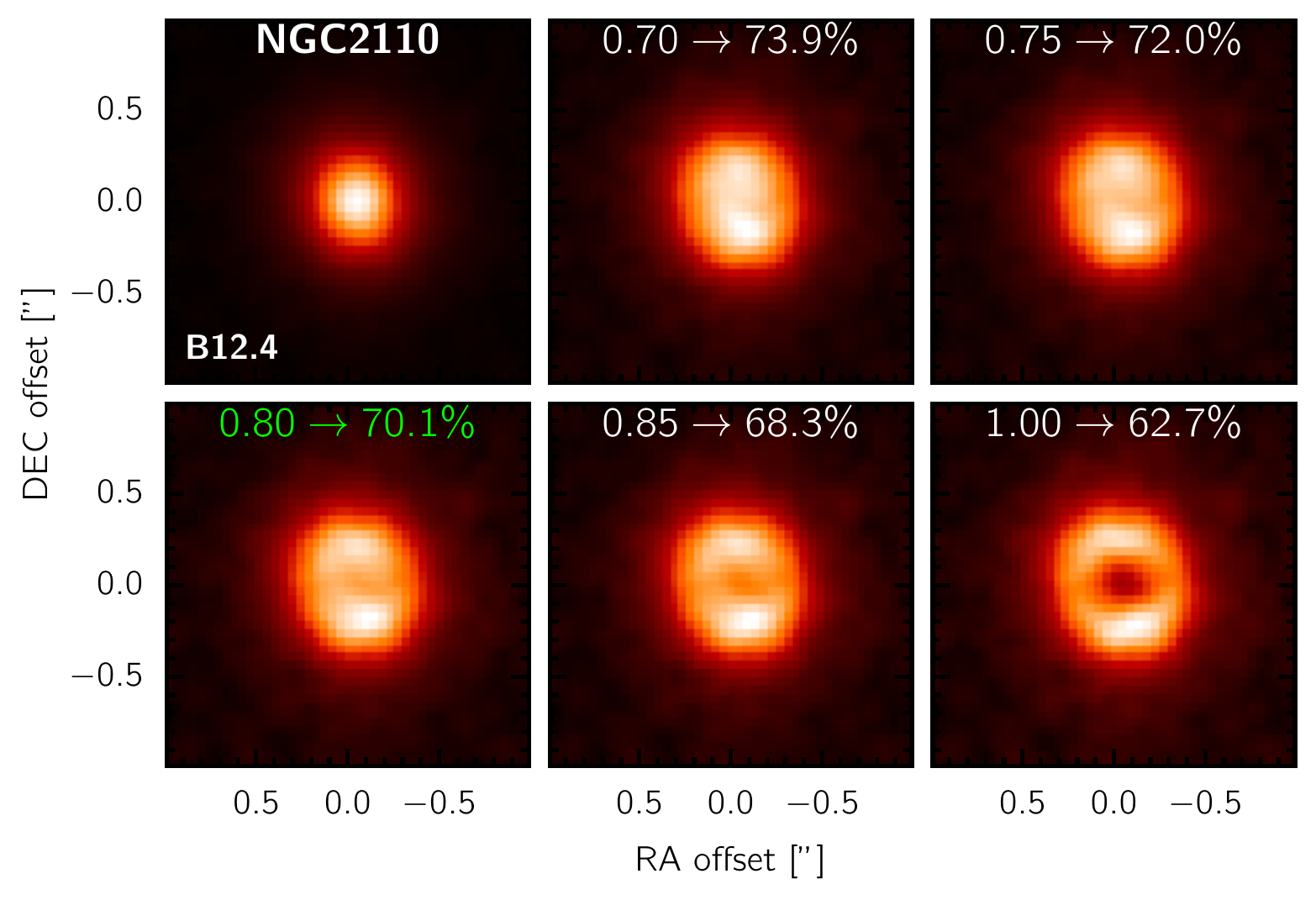}  
            \includegraphics[angle=0,width=1.0\columnwidth]{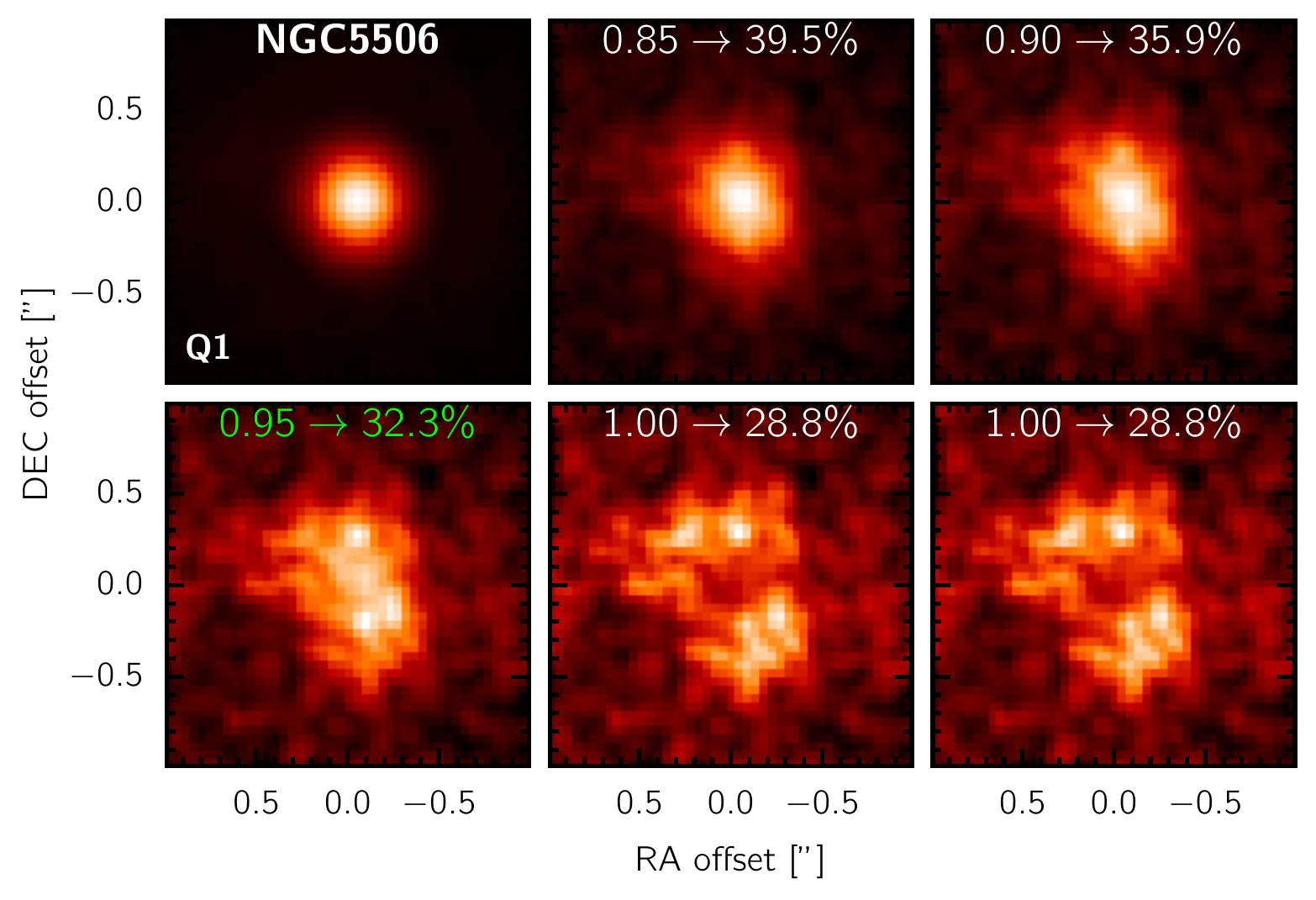}  
                  \includegraphics[angle=0,width=1.0\columnwidth]{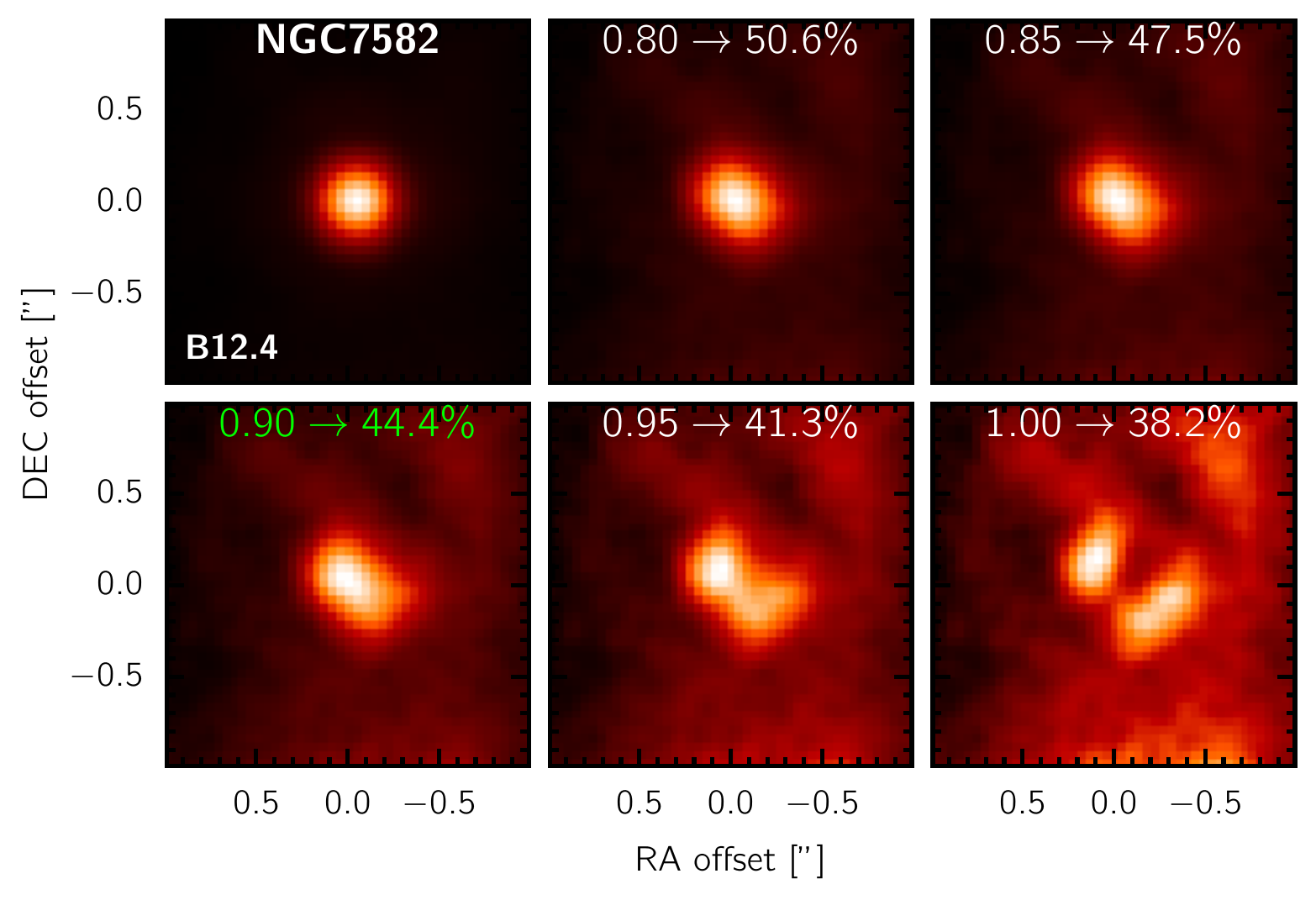}        
    \caption{
             Same as Fig.~\ref{fig:sub} but for the other objects from the sample, except 3C\,321 and NGC\,5135.
             }
\end{figure*}

\end{document}